\documentclass[12pt]{article}
\usepackage{mathrsfs}
\usepackage{cite}
\usepackage{hyperref}
\hypersetup{colorlinks=true,backref=true,linkcolor=black,anchorcolor=black,citecolor=black,filecolor   =black,menucolor=black,pagecolor=black,urlcolor=black}
\usepackage[english]{babel}
\newlength{\abstractwidth}
\usepackage{epsf}
\usepackage{color}
\usepackage{graphicx}
\usepackage{amssymb}
\usepackage{latexsym}

\usepackage{color}
\definecolor{dgreen}{rgb}{0,0.70,0.30}
\definecolor{gold}{rgb}{0.85,.66,0}
\definecolor{purple}{rgb}{1.0,0.3,0.6}
\definecolor{red}{rgb}{1.0,0.0,0.0}

\usepackage{tikz}
\usetikzlibrary{calc,decorations.pathreplacing}

\def\ba{\begin{align}}
\def\ea{\end{align}}
\def\bse{\begin{subequations}}
\def\ese{\end{subequations}}
\def\Im{\,{\rm Im}\,}
\def\Re{\,{\rm Re}\,}

\def\cA{{\cal A}}
\def\cB{{\cal B}}
\def\cC{{\cal C}}
\def\cD{{\cal D}}
\def\cE{{\cal E}}
\def\cF{{\cal F}}
\def\cG{{\cal G}}

\def\cM{{\cal M}}

\def\cO{{\cal O}}

\def\cR{{\cal R}}

\def\ba{{\bf a}}

\def\ttau{{\tilde \tau}}
\def\ty{{\tilde y}}
\def\tx{{\tilde x}}

\def\mH{\mathfrak{H}}

\def\p{\partial}
\def\half{{1\over 2}}

\def\f{\varphi}
\def\no{\nonumber}
\def\sm{\smallskip}

\def\NN{{\mathbb N}} 
\def\RR{{\mathbb R}}

\def\ZZ{{\mathbb Z}}

\def\tE{E}

\def\half{ {1\over 2}}

\renewcommand{\thefootnote}{\fnsymbol{footnote}}
\renewcommand{\thanks}[1]{\footnote{#1}}
\newcommand{\starttext}{
\setcounter{footnote}{0}
\renewcommand{\thefootnote}{\arabic{footnote}}}

\newcommand{\bea}{\begin{eqnarray}}
\newcommand{\eea}{\end{eqnarray}}
\newcommand{\be}{\begin{eqnarray}}
\newcommand{\ee}{\end{eqnarray}}

\begin{document}
\starttext
\setcounter{footnote}{0}

\begin{flushright}
2015 October 12 \\
DAMTP-2015-48 \\
IPhT-t15/134 \\
IHES/P/15/07  
\end{flushright}

\bigskip

\begin{center}

{\Large \bf Proof of a modular  relation between 1-, 2- and 3-loop }

\vskip 0.05in 

{\Large \bf Feynman diagrams on a torus}

\vskip 1cm

{\large \bf Eric D'Hoker$^{(a)}$, Michael B. Green$^{(b)}$ and Pierre Vanhove$^{(c)}$}
\vskip 0.2cm

{ \sl (a) Department of Physics and Astronomy }\\
{\sl University of California, Los Angeles, CA 90095, USA} 

\vskip 0.08in

{ \sl (b) Department of Applied Mathematics and Theoretical Physics }\\
{\sl Wilberforce Road, Cambridge CB3 0WA, UK}

\vskip 0.08in

{\sl (c) Institut des Hautes \'Etudes Scientifiques\\
 Le Bois-Marie, 35 route de Chartres,
 F-91440 Bures-sur-Yvette, France, \\
 and \\
Institut de physique th\'eorique,\\
 Universit\'e Paris Saclay, CEA, CNRS, F-91191 Gif-sur-Yvette}

\vskip 0.1in

{\tt \small dhoker@physics.ucla.edu; M.B.Green@damtp.cam.ac.uk; pierre.vanhove@cea.fr}

\end{center}

\vskip -1 in

\begin{abstract}
The coefficients of the higher-derivative terms in the low energy expansion of genus-one graviton Type II superstring scattering amplitudes are determined by integrating sums of non-holomorphic modular functions over the complex structure modulus of a torus.  In the case of the four-graviton amplitude, each of these modular functions is a multiple sum associated with a Feynman diagram for a  free massless scalar field on the torus. The lines in each diagram join pairs of vertex insertion points and the number of lines defines its weight $w$, which corresponds to  its order in the low energy expansion.  Previous results concerning the low energy expansion of the genus-one four-graviton amplitude led to a number of conjectured relations between modular functions of a given $w$, but different numbers of loops  $\le w-1$.  In this paper we shall prove the simplest of these conjectured relations, namely the one that arises at weight $w=4$ and expresses the three-loop modular function $D_4$ in terms of modular functions with one and two loops. As a byproduct, we prove three intriguing new holomorphic modular identities.

  \end{abstract}

\newpage
\setcounter{tocdepth}{1}

\newpage

\section{Introduction and summary of results}
\setcounter{equation}{0}
\label{intro}

In an earlier paper \cite{D'Hoker:2015foa} we elucidated certain properties of the non-holomorphic modular functions that enter into the low energy expansion of the genus-one four-graviton scattering amplitude in Type II superstring perturbation theory.\footnote{Non-holomorphic modular functions are  functions of the complex structure parameter~$\tau$ of the torus and its complex conjugate~$\bar \tau$ which are invariant under the canonical action of $SL(2,\ZZ)$ on~$\tau$ and~$\bar \tau$ by M\"obius transformations. For paedagigical introductions to the subject, one may consult \cite{Zagier:2008} \cite{Goldfeld}, \cite{Terras}, while a historical perspective is presented in \cite{Weil:1977}.
Since we shall deal with both holomorphic and non-holomorphic functions of~$\tau$ in this paper, we shall indicate the dependence on both~$\tau$ and~$\bar \tau$ for non-holomorphic objects throughout, which is a different notation from that adopted in \cite{D'Hoker:2015foa}.} These non-holomorphic modular functions arise from vacuum Feynman diagrams for a massless scalar field on a torus of fixed modulus~$\tau$ with marked points at the positions of the four vertex operators.  The lines in a diagram correspond to Green functions (i.e., propagators) joining pairs of these points.   The {\sl weight }~$w$ is the number of scalar lines in the diagram; it governs the order at which the diagram contributes to the low energy expansion. The number of loops of a diagram will be denoted $L$. Expressing a Feynman diagram in terms of the discrete momenta on the torus gives a representation of its value in terms of a  multiple sum over $2L$ independent integers that are generalizations of the standard non-holomorphic Eisenstein series (for which $L=1$)\footnote{These bear some resemblance to multiple Kronecker--Eisenstein series of the type discussed in  \cite{Goncharov:2008}.}. There are contributions from diagrams of weight $w$  with different numbers of loops $L$, subject to the constraint $L \leq w-1$.  A key to the progress made in elucidating the properties of these modular functions in \cite{D'Hoker:2015foa}
was understanding the structure that emerges by considering families of modular functions with a fixed number of loops $L$. 

\sm

$\bullet$ For $L=1$ (which is the lowest non-trivial value for $L$ due to momentum conservation on the torus) and weight $w$, the modular function is unique (up to a constant normalization factor) and given by the non-holomorphic Eisenstein series $E_w$, defined by,
\bea
\label{1a1}
E_w (\tau, \bar \tau) = \sum _{(m,n)\not= (0,0)}  \left ( { \tau_2 \over \pi \, |m\tau + n|^2} \right )^w
\eea 
The sum is over integers $m,n \in \ZZ$ which parametrize the discrete momenta on the torus; 
the real and imaginary parts of $\tau$ are respectively $\tau_1, \tau_2 \in \RR$; and the factor of $1/\pi ^w$ has been included for convenience. The Eisenstein series satisfies the Laplace-eigenvalue equation,
\bea
\label{1a2}
\Delta E_w = w(w-1) E_w
\eea
where the Laplace-Beltrami operator $\Delta$ on the upper half plane is given by $\Delta = 4 \tau _2 ^2 \, \p_\tau \p_{\bar \tau}$.

\sm

$\bullet$ For $L=2$, the most general vacuum Feynman diagram of weight $w$ is given by a multiple sum  of the form, 
\bea
\label{1a3}
C_{a_1,a_2, a_3} (\tau, \bar \tau) = \sum _{(m_r, n_r) \not= (0,0)}  \delta _{m,0} \, \delta _{n,0} \, 
\prod _{r=1}^3 \left ( {  \tau _2 \over \pi  \, |m_r \tau+n_r|^2} \right )^{a_r} 
\eea
The sum is over integers $m_r , n_r \in \ZZ$ for $r=1,2,3$, while $m = m_1+m_2+m_3$ and $n=n_1+n_2+n_3$ are constrained to vanish by the Kronecker symbols. The parameters $a_1, a_2, a_3$ are positive integers subject to $w=a_1+a_2+a_3$. A rich structure of the space of all $L=2$ modular functions of arbitrary given weight $w$ was exhibited in \cite{D'Hoker:2015foa}. This was achieved by showing that the functions $C_{a_1, a_2, a_3}$ satisfy a system of Laplace eigenvalue equations whose inhomogeneous parts are quadratic polynomials in Eisenstein series, and whose eigenvalues and  eigenspaces are governed by the representation theory of a hidden $SO(2,1)$.
The simplest examples may be exhibited for the lowest weights $w=3,4$ where we have,
\bea
\label{1a4}
\Delta C_{1,1,1} & = & 6 \, E_3
\no \\
(\Delta -2) C_{2,1,1} & = & 9 \, E_4 - E_2^2
\eea
The first equation may be integrated and the integration constant may be determined by 
matching the asymptotic behavior near the cusp $\tau _ 2 \to \infty$ to give $C_{1,1,1}= E_3 + \zeta (3)$, a result that had been obtained earlier by Zagier by direct summation of the series \cite{Zagier:2014}. The second equation admits no such simple integration, but its significance will become clear shortly.

\sm

$\bullet $ For $L \geq 3$, the situation is more complicated and considerably less well-understood. There is no longer a single formula (such as for $C_{a,b,c}$ for $L=2$)  to evaluate all diagrams, since more than a single diagram topology contributes when $L\geq 3$.  Moreover, there is no systematic way known to derive equations of the Laplace eigenvalue type  for the corresponding  multiple sums. Therefore, the methods used to expose the structure at two loops appear of little use for higher loop diagrams. It has been possible, however, to formulate certain conjectured relations between the weight four and weight five modular functions. 

\sm

The simplest of these conjectures was for the modular function $D_4$ 
characterized by  three loops, $L=3$, and weight $w=4$, and given by the following sum,
\bea
\label{1a5}
D_4 ( \tau , \bar \tau) = \sum _{ (m_r,n_r)\not=0 } 
 \delta _{m,0} \, \delta _{n,0} \, \prod _{r=1}^4  \left ( { \tau_2 \over \pi \, |m_r  \tau +n_r|^2  }  \right )
 \eea
The sum is over integers $m_r, n_r \in \ZZ$ with $r=1,2,3,4$, subject to the vanishing of $m=m_1+m_2+m_3+m_4$ and $n= n_1+n_2+n_3+n_4$ as enforced by the Kronecker $\delta$-symbols.

\sm 

The purpose of the present paper is to provide an analytical proof of  the conjecture for $D_4$,
which we shall henceforth refer to as the Theorem for $D_4$. This involves the following polynomial combination   of modular functions,
\bea
\label{1a6}
F(\tau , \bar \tau) = D_4(\tau , \bar \tau) -24 C_{2,1,1}(\tau , \bar \tau) -3 E_2(\tau , \bar \tau)^2 + 18 E_4 (\tau , \bar \tau)
\eea
By construction, $F$ is a modular function of weight 4 and  involves contributions with one, two, and three loops.
The precise statement is as follows.

\sm\sm 
 
{\bf Theorem}:\ {\sl The modular function $D_4( \tau , \bar \tau)$ satisfies the relation}
\bea
\label{1a7}
F(\tau , \bar \tau) =0   
\eea
The evidence for the validity of (\ref{1a7}) presented in \cite{D'Hoker:2015foa} was based on an analysis of the behavior of $F(\tau , \bar \tau)$ near the cusp $\tau_2 \to \infty$. The  expansion of $F(\tau , \bar \tau)$ in powers of $q$ and $\bar q$,
whose coefficients are powers of $\tau_2$, was verified to vanish to lowest and first orders in $q $ and $\bar q$.
This gave us compelling evidence but, of course, did not constitute an analytic proof of (\ref{1a7}). 
To prove the Theorem, we shall first prove the following auxiliary Lemma.

\sm\sm 
 
{\bf Lemma}:  
\begin{enumerate}
\item {\sl The modular function $F( \tau , \bar \tau)$, defined in (\ref{1a6}), admits a decomposition in inverse powers of $\tau_2$, with a finite number of terms,
\bea
\label{1a8}
F(\tau , \bar \tau) = \sum _{k=0} ^7 (\pi \tau_2)^{4-k} \cF_k (q, \bar q)
\eea
where the coefficients $\cF_k (q, \bar q)$ are entire functions of $q=e^{2 \pi i \tau}$ and $\bar q = e^{-2 \pi i \bar \tau}$;}
\item {\sl The coefficients $\cF_k$ vanish for $k=0,1,2,3,7$;}
\item {\sl The coefficients $\cF_k$ for $k=4,5,6$  are harmonic functions of $q$ and $\bar q$, 
which may be expressed in terms of functions $\f_k(q)$ which are holomorphic in the unit disc $|q|<1$,
\bea
\label{1a9}
\cF_k (q,\bar q) = \f_k (q) + \overline{\f_k (q)}
\eea
where $\f_k(q)$ obeys the conjugation properties,
\bea
\label{1a10}
\overline{\f_k(q)} = \f_k (\bar q) \hskip 1in \f_k (1/q) = (-)^k \f _k (q)
\eea
The first relation in (\ref{1a10}) is complex conjugation and implies that the Taylor series of $\f_k(q)$ in 
powers of $q$ has real coefficients.}
\end{enumerate}
The proof of the Lemma will proceed by direct summation over the integers $n_r$ in the  multiple sum that defines $D_4$ in (\ref{1a5}), $C_{2,1,1}$ in (\ref{1a3}), and $E_2, E_4$ in (\ref{1a1}).
This method generalizes the calculation of Zagier for the function $C_{1,1,1}$. The proof of items 1 and 2 will be relatively straightforward, but the proof of item 3 will require extensive algebraic manipulations, most of which will be summarized in Appendix \ref{appB}. 

\sm

The proof of the Theorem itself will proceed by showing that the holomorphic functions $\f_k(q)$ for $k=4,5,6$
are modular forms of weight $0, -2$, and $-4$ respectively, and therefore must vanish ($\f_4$ must be constant by this argument, but will vanish in view of its asymptotic behavior near the cusp). After the first version of this paper was submitted to the arXiv,  Pierre Deligne sent us a more direct derivation of the vanishing of $\f_k(q)$ for $k=4,5,6$, which we include in appendix \ref{appD}.

\sm\sm

{\bf Holomorphic Corollaries}:  

\sm

The vanishing of the holomorphic coefficients $\f_k(q)$ for $k=4,5,6$ leads to three holomorphic 
identities given in (\ref{B7}), (\ref{B22}) and (\ref{B31}).  While their vanishing is a consequence 
of the Theorem, no direct analytical proof of the identities is known to us. The simplest of these identities results from 
splitting $\f_6(q) = \f_6 ^{(1)} (q) + \f_6 ^{(2)}(q)$ and then proving by combinatorial 
rearrangements that $\f_6 ^{(1)}=0$, which leaves the non-trivial identity $\f_6 ^{(2)}(q)=0$, namely,
\bea
\label{1a11}
 {15 \over 8}  \zeta (6) & = &
 \sum _{m_1+m_2 \not= 0} ' { 3   \over 16 m_1 m_2 (m_1+m_2)^4} 
  { (1+q^{m_1}) (1 + q^{m_2}) \over (1-q^{m_1}) (1-q^{m_2}) } 
\no \\ &&  
+ \sum _{m_1,m_2}' { 3  \over 16 m_1^3 m_2^3} \, 
{ (1+q^{m_1})(1+q^{m_2})  \over (1- q^{m_1})(1-q^{m_2}) } 
 - \sum _{m} ' { 9 \over 4 m^6 } \, { q^m \over ( 1- q^m)^2} 
\eea
The sum extends over integers $m_1, m_2 \in \ZZ$ which do not vanish (as indicated by the prime). 
A direct combinatorial proof of this identity is not known to us, but we have confirmed its validity  
order by order in an expansion in $q$ around $q=0$ using MAPLE to order $\cO(q^{400})$.
 
 \sm
 
Our  proof of  the $D_4$ conjecture of  \cite{D'Hoker:2015foa} provides significant encouragement that the conjectures 
advanced in \cite{D'Hoker:2015foa} for the three non-trivial weight $w=5$ modular functions,
namely $D_5, D_{3,1,1}$ and $D_{2,2,1}$ may be proven by the same methods, even if the algebraic manipulations involved will be even more arduous. One would hope that with more insight a simpler proof will emerge, 
which would facilitate the proofs for  $D_5, D_{3,1,1}$ and $D_{2,2,1}$ and the generalizations to higher weight. We shall report on such developments in future work.

\subsection{Organization}

The remainder of this paper is organized as follows. 
In section \ref{sec2} we present a brief review of the role
of non-holomorphic modular functions in the low energy expansion of genus-one string perturbation theory.
In section \ref{sec3}, we provide a proof of the Lemma, supported by results derived in Appendices \ref{appA}
and \ref{appB}. In section \ref{sec4}, we provide a proof of the Theorem by combining the results of the Lemma and further results on the structure of $F(\tau, \bar \tau)$ derived in Appendix \ref{appC}. An alternative derivation by  Deligne of the results of section \ref{sec4} is given in Appendix D. In section \ref{sec5}, we spell out the various components of the Holomorphic Corollaries, and exhibit the holomorphic identities which arise as spin-offs of the proof of the Theorem.  Finally, in section \ref{sec6} we summarize our results, and discuss the outlook of our work.

\section{String theory origin of the modular functions}
\setcounter{equation}{0}
\label{sec2}

We will here give a brief overview of the motivation for considering non-holomorphic modular functions from the physical perspective of superstring perturbation theory.  This section is not essential for the remainder of the paper, whose purpose is rather to give a purely mathematical proof of the Theorem stated in the introduction. The starting point is the full genus-one four-graviton amplitude $\cA_1$ in Type II closed superstring perturbation theory. It is given by an integral over the moduli space of genus-one Riemann surfaces of a partial amplitude $\cB_1$ which is defined at fixed modulus.  
It is this partial amplitude $\cB_1$ that will be of direct interest in this paper, and we shall start by reviewing its structure. For completeness, we shall also include a brief discussion of the structure of the full amplitude $\cA_1$. 
 
\subsection{The partial amplitude $\cB_1$ at fixed modulus}
 
We consider a torus with  modulus $\tau \in \mH$ where $\mH= \{\tau = \tau_1 + i \tau_2, \, \tau_1, \tau_2 \in \RR, \, \tau_2>0\}$.
The partial amplitude $\cB_1$ is a family of non-holomorphic modular functions $\cB_1 (s,t,u|\tau, \bar \tau)$ which may be defined in terms of an exponential of the scalar Green function $G$ on the torus,
\bea
\label{2a1}
\cB_1 (s,t,u | \tau, \bar \tau) = 
\prod _{i=1}^4 \int _\Sigma {d^2 z_i \over \tau_2} \,  \exp \left \{ \sum _{1 \leq j<k \leq 4} s_{jk}  \, G(z_j-z_k |  \tau, \bar \tau )  \right \}
\eea
The integral is over four copies of the Riemann surface $\Sigma$ of modulus $\tau$.  
The parameters $s_{12}=s_{34}=s$, $s_{23}=s_{14}=t$, and $s_{13}=s_{24} =u$ are dimensionless Lorentz invariants $s_{ij}=-\alpha ' k_i \cdot k_j/2$ of the momenta $k_i$ of the four gravitons labelled by $i,j=1,2,3,4$. They obey the relation $s+t+u=0$ in view of overall momentum conservation for massless states. We exhibit all three parameters $s,t,u$ -- despite their interdependence -- because $\cB_1 (s,t,u|\tau, \bar \tau)$ is in fact a symmetric function of $s,t,u$ due to Bose symmetry of the gravitons.

\sm

The scalar Green function $G(z |  \tau, \bar \tau  )$ on the torus satisfies $ \p_z \p_{\bar z} G(z | \tau, \bar \tau) = \pi \delta ^{(2)} (z) -\pi /\tau_2$. In view of the relation $s+t+u=0$, the Green function $G$  may be shifted by an arbitrary  $z$-independent quantity without affecting the  integrand in (\ref{2a1}), or of $\cB_1$ itself. We use this symmetry to impose the normalization condition $\int _\Sigma d^2z \, G(z| \tau, \bar \tau)=0$ on $G$, and express the resulting Green function as a Fourier sum over the integers $m,n \in \ZZ$,
\bea 
\label{2c3}
G (z| \tau , \bar \tau) =  \sum_{(m,n) \neq(0,0)} 
\cG (m,n |  \tau , \bar \tau ) \, e^{2\pi  i (m \alpha   - n \beta)}   
\eea 
where $\alpha, \beta \in \RR$ parametrize $z$ by $z=\alpha + \beta \tau$.
The Fourier modes $\cG(m,n|  \tau , \bar \tau )$ are given by,
\bea
\cG (m,n | \tau , \bar \tau ) = { \tau_2\over \pi  |m \tau +n|^2}    
\label{propdef}
\eea
and we shall set $\cG(0,0| \tau, \bar \tau)=0$ by convention. The integers $m,n$ label the two-dimensional momenta of the scalar field on the torus, the zero mode being excluded. For fixed $\tau$, $G(z |\tau, \bar \tau) $ is regular in $z$,
except for a logarithmic singularity at the origin where $G(z|\tau, \bar \tau) \sim - \ln |z|^2 $.

\subsection{The low energy expansion} 

For fixed $\tau$, the singularities of $\cB_1$ as a function of $s,t,u$  are simple and double poles located at positive integer values of $s, t, u$. Low energy corresponds to $|s|, \, |t|, \, |u| \ll 1$, which is a region where $\cB_1$ is analytic and admits a Taylor series expansion in $s,t,u$ with finite radius of convergence.  This expansion  was investigated in \cite{D'Hoker:2015foa}, following less complete work in  \cite {Green:1999pv, Green:2008uj}. The low energy expansion may be obtained by Taylor expanding the exponential of (\ref{2a1}) in its argument, and denoting the total degree of homogeneity in the variables $s,t,u$ by the weight $w$, 
\bea
\label{2b5}
\cB_1 (s,t,u |  \tau , \bar \tau)
=  \sum _{w=0} ^{\infty} { 1 \over w!} \, 
\prod _{i=1}^4 \int _\Sigma {d^2 z_i \over \tau_2} \,  \left ( \sum _{1 \leq j<k \leq 4} s_{jk}  \, G(z_j-z_k |  \tau, \bar \tau )  \right )^w
\eea  
The coefficients in this expansion are modular functions of $ \tau $.  
They can be represented by sums of vacuum Feynman diagrams in which the four vertices labelled by $i=1,2,3,4$ represent the integration points $z_i$,  and in which $\ell_{jk}$ lines represent the Green function of  (\ref{2c3}) joining the vertices $j$ and $k$.  The weight of the diagram is given by, 
\be
\label{weightdiag}
w= \sum_{1\le j<k \le 4}^6 \ell_{jk}   
\ee    
Carrying out the integrations over the vertex positions $z_i$ will then produce the multiple sums that were described in the introduction. A systematic discussion of the combinatorial notation for general Feynman diagrams encountered in the low energy expansion of $\cB_1$ was presented in \cite{D'Hoker:2015foa} following  \cite {Green:1999pv, Green:2008uj}, but will not be needed here. Instead, we shall content ourselves with writing down the Feynman diagrams that correspond to the limited number of functions needed in this paper.

\sm

We denote the Green function $G(z_j-z_k|\tau, \bar \tau)$ by a  line between the vertices $i$ and $j$,
\begin{center}
\begin{tikzpicture} [scale=1.3]
\scope[xshift=0cm,yshift=0cm]
\draw (-1.5,0) node{$G(z_j-z_k|\tau, \bar \tau) \, = $} ;
\draw (0,-0.0) node{$\bullet$}   ;
\draw (1,0.0) node{$\bullet$}  ;
\draw (0,0) -- (1,0) ;
\draw (0,0.3) node{$j$} ;
\draw (1,0.3) node{$k$} ;
\endscope
\end{tikzpicture}
\end{center} 
We shall represent an integrated vertex  by an unlabeled dot. 
For example, the integrated string of two Green functions is denoted as follows,
\begin{center}
\begin{tikzpicture} [scale=1.3]
\scope[xshift=0cm,yshift=0cm]
\draw (-3.1,0) node{$\int _\Sigma { d^2z_i \over \tau_2} \, G(z_j-z_i|\tau, \bar \tau) \, G(z_i-z_k|\tau, \bar \tau) \, = $} ;
\draw (0,-0.0) node{$\bullet$}   ;
\draw (1,0.0) node{$\bullet$}  ;
\draw (2,0.0) node{$\bullet$}  ;
\draw (0,0) -- (2,0) ;
\draw (0,0.3) node{$j$} ;
\draw (2,0.3) node{$k$} ;
\endscope
\end{tikzpicture}
\end{center}  
For an integrated string with  $a$ Green functions, we shall use the following notation, 
\begin{center}
\begin{tikzpicture} [scale=1.3]
\scope[xshift=0cm,yshift=0cm]
\draw (0,0) node{$\bullet$}   ;
\draw (1,0) node{$\bullet$}  ;
\draw (2,0) node{$\bullet$}   ;
\draw (3,0) node{$\bullet$}  ;
\draw (4,0) node{$\bullet$}  ;
\draw (5,0) node{$\bullet$}  ;
\draw (6,0) node{$\bullet$}  ;
\draw (0,0) -- (1,0) ;
\draw (2,0) -- (4,0) ;
\draw[dashed, thick] (4,0) -- (5,0) ;
\draw (5,0) -- (6,0) ;
\draw (0.65,-0.17) [fill=white] rectangle(0.33cm, 0.15cm) ;
\draw (0.5,0.0) node{$a$};
\draw (0,0.3) node{$j$} ;
\draw (1,0.3) node{$k$} ;
\draw (1.5,0) node{$=$} ;
\draw (2,0.3) node{$j$} ;
\draw (6,0.3) node{$k$} ;
\endscope
\end{tikzpicture}
\end{center}   
Given this notation, it is straightforward to express the multiple sums encountered in the introduction
in terms of Feynman diagrams. For $L=1$, we have,
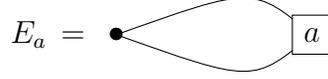
\begin{center}
\begin{tikzpicture} [scale=1.3]
\scope[xshift=0cm,yshift=0cm]
\draw (-0.7,0) node{$E_a \, =$} ;
\draw (0,0) node{$\bullet$} ..controls (1.4,0.5) ..  (2,0) node{$\bullet$};
\draw (0,0) node{$\bullet$} ..controls (1.4,-0.5) ..  (2,0) node{$\bullet$};
\draw (2.2,-0.2) [fill=white] rectangle(1.8,0.2) ;
\draw (2,0.0) node{$a$};
\endscope
\end{tikzpicture}
\end{center}   
where the vertex on the left side of the diagram is to be integrated, as is consistent with the notation of the unlabelled dot. For $L=2$, we have, 
 
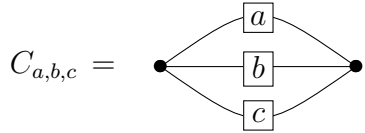
\begin{center}
\begin{tikzpicture} [scale=1.3]
\scope[xshift=0cm,yshift=0cm]
\draw (-1,0) node{$C_{a,b,c}  \, =$} ;
\draw (0,0) node{$\bullet$} ..controls (1,0.7) ..  (2,0) node{$\bullet$};
\draw (0,0) -- (2,0) ;
\draw (0,0) node{$\bullet$} ..controls (1,-0.7) ..  (2,0) node{$\bullet$};
\draw (1.15,0.35) [fill=white] rectangle(0.85, 0.65) ;
\draw (1,0.5) node{$a$};
\draw (1.15,-0.2) [fill=white] rectangle(0.85, 0.15) ;
\draw (1,0.0) node{$b$};
\draw (1.15,-0.65) [fill=white] rectangle(0.85, -0.35) ;
\draw (1.0, -0.5) node{$c$};
\endscope
\end{tikzpicture}
\end{center}   
The lowest weight example with  $L=3$ is the modular function $D_4$ of weight $w=4$, which is  the function of central interest in this paper and is associated with the following diagram, 
 
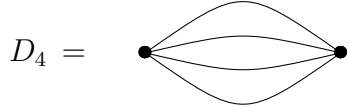
\begin{center}
\begin{tikzpicture} [scale=1.3]
\scope[xshift=0cm,yshift=0cm]
\draw (-1,0) node{$D_4  \, =$} ;
\draw (0,0) node{$\bullet$} ..controls (1,0.7) ..  (2,0) node{$\bullet$};
\draw (0,0) node{$\bullet$} ..controls (1,-0.7) ..  (2,0) node{$\bullet$};
\draw (0,0) node{$\bullet$} ..controls (1,0.23) ..  (2,0) node{$\bullet$};
\draw (0,0) node{$\bullet$} ..controls (1,-0.23) ..  (2,0) node{$\bullet$};
\endscope
\end{tikzpicture}
\end{center}    
Finally, we also list the modular functions of weight 5 that were also the subject of conjectures in
 \cite{D'Hoker:2015foa} that will be described later,
 
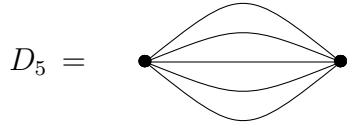
\begin{center}
\begin{tikzpicture} [scale=1.3]
\scope[xshift=0cm,yshift=0cm]
\draw (-1,0) node{$D_5  \, =$} ;
\draw (0,0) node{$\bullet$} ..controls (1,0.8) ..  (2,0) node{$\bullet$};
\draw (0,0) node{$\bullet$} ..controls (1,-0.8) ..  (2,0) node{$\bullet$};
\draw (0,0) node{$\bullet$} --  (2,0) node{$\bullet$};
\draw (0,0) node{$\bullet$} ..controls (1,0.4) ..  (2,0) node{$\bullet$};
\draw (0,0) node{$\bullet$} ..controls (1,-0.4) ..  (2,0) node{$\bullet$};
\endscope
\end{tikzpicture}
\end{center}     
 
  
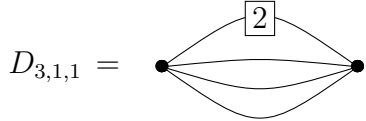
\begin{center}
\begin{tikzpicture} [scale=1.3]
\scope[xshift=0cm,yshift=0cm]
\draw (-1,0) node{$D_{3,1,1}  \, =$} ;
\draw (0,0) node{$\bullet$} ..controls (1,0.7) ..  (2,0) node{$\bullet$};
\draw (0,0) node{$\bullet$} ..controls (1,-0.7) ..  (2,0) node{$\bullet$};
\draw (0,0) node{$\bullet$} ..controls (1,0.1) ..  (2,0) node{$\bullet$};
\draw (0,0) node{$\bullet$} ..controls (1,-0.3) ..  (2,0) node{$\bullet$};
\draw (1.15,0.32) [fill=white] rectangle(0.85, 0.67) ;
\draw (1,0.5) node{$2$};
\endscope
\end{tikzpicture}
\end{center}   
 
 
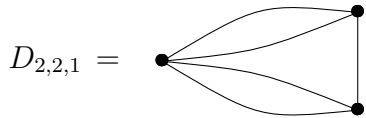
\begin{center}
\begin{tikzpicture} [scale=1.3]
\scope[xshift=0cm,yshift=0cm]
\draw (-1,0) node{$D_{2,2,1}  \, =$} ;
\draw (0,0) node{$\bullet$} ..controls (1,0.6) ..  (2,0.5) node{$\bullet$};
\draw (0,0) node{$\bullet$} ..controls (1,-0.6) ..  (2,-0.5) node{$\bullet$};
\draw (2,0.5) node{$\bullet$} --  (2,-0.5) node{$\bullet$};
\draw (0,0) node{$\bullet$} ..controls (1,0.1) ..  (2,0.5) node{$\bullet$};
\draw (0,0) node{$\bullet$} ..controls (1,-0.1) ..  (2,-0.5) node{$\bullet$};
\endscope
\end{tikzpicture}
\end{center}     
The modular function $D_5$ has $L=4$, while $D_{3,1,1}$ and $D_{2,2,1}$ have $L=3$. 
The expressions for these modular functions in terms of multiple sums are easily obtained from the diagrams.   For example, for $D_{2,2,1}$, we have,
\bea
D_{2,2,1} = \sum _{(m,n)} ^\prime  {\tau_2 \over \pi \, |m+n \tau|^2} 
\prod _{i=1}^2 \Bigg ( \sum _{(m_i,n_i)} ^\prime { \tau_2 ^2 \over  \pi^2 |m_i+n_i \tau|^2 \,  |m+m_i+(n+n_i)\tau |^2 } \Bigg )
\eea
where the prime superscripts on the sums indicate that the zero mode is to be omitted.

 \subsection{The conjectured relations at weight 5}
 
In addition to the conjectured relation for $D_4$ stated in the Theorem (\ref{1a7}) of the introduction,
and which is to be proven in this paper, a number of further relations for weight five modular functions 
were conjectured in  \cite{D'Hoker:2015foa}.   These comprise  the following relations,
  \bea
 \label{D5}
 0 &=& D_5 -60 \, C_{3,1,1} - 10  \, \tE_2 \, C_{1,1,1} +48 \, \tE_5 -16 \, \zeta(5)    
 \no \\
0 & = & 40 \, D_{3,1,1}-300 \, C_{3,1,1} - 120 \,  \tE_2 \, \tE_3 +276 \, \tE_5    - 7 \, \zeta(5)  
\no \\
0 & = & 10 \, D_{2,2,1} - 20 \, C_{3,1,1}+ 4  \, \tE_5 - 3 \, \zeta(5)   
 \eea
Here, $\zeta(w)$ denotes the Riemann $\zeta$-function, to which we assign weight $w$. Each equation in (\ref{D5}) relates a weight five $D$-function to a weight five polynomial in modular functions of lower loop number, i.e. lower depth.  We expect that the conjectures of (\ref{D5}) may be proven as well by the methods used in this paper. Further relations should be expected to proliferate at higher weights $w$ and it would be fascinating to understand their complete structure.

\subsection{The full amplitude as an integral over moduli}

The full genus-one four-graviton amplitude  in Type II superstring theories, $\cA_1 $,  is obtained by integrating the partial amplitude $\cB_1$ over the moduli space $\cM_1$ of genus-one Riemann surfaces \cite{Green:1981yb}, 
\bea
\cA_1 = \kappa ^2 \,  \cR^4\, \int_{\cM_1} d\mu_1\, \cB_1(s,t,u| \tau , \bar \tau)   
\label{amp4}
\eea
The symbol $\cR^4$ denotes four powers of the linearized Riemann curvature tensor,  and the normalization $\kappa ^2$  is proportional to Newton's constant in ten-dimensional space-time.   The integral is over a fundamental domain $\cM_1$ of the modular group $SL(2,\ZZ)$ acting on the upper half plane $\mH$,  and  $d \mu_1=  d  \tau_1\, d  \tau_2/ \tau_2^2$ is the volume form of the Poincar\'e metric.  

\sm

The integral of $\cB_1$ over $\cM_1$ is absolutely convergent only for purely imaginary $s,t, u$. Constructing and evaluating $\cA_1$ beyond this region requires analytic continuation. This analytic continuation was shown to exist, 
to be computable, and to  produce  dependences on $s,t,u$ which are no longer analytic near the origin $s=t=u=0$
in \cite{D'Hoker:1994yr}. The physical origin for this non-analytic behavior is well-known and well-understood, 
as it results from the propagation of massless string states in closed loops.

\sm

The emergence of non-analyticities at low energy means that the expansion in powers of $s,t,u$ in (\ref{2b5}) 
and the integration of $\cB_1$ over moduli space $\cM_1$ in (\ref{amp4}) cannot be interchanged, 
since doing so would produce divergent integrals. To evaluate the low energy expansion, 
one may first extract the exact non-analytic behavior to a given order in $s,t,u$, 
and then evaluate the remaining finite part which is polynomial in $s,t,u$. 
Alternatively, space-time may be partially compactified on a flat torus $T^d$, 
and the non-analytic part may be canceled when comparing the low energy 
contributions at different moduli of $T^d$. We refer the reader to \cite{D'Hoker:2015foa}
for detailed discussions of both approaches.

\section{Proof of the Lemma}
\setcounter{equation}{0}
\label{sec3}
 
In this section, we shall provide a proof of the Lemma  given in (\ref{1a8})-(\ref{1a10}).

\sm

The strategy for proving part 1 of the Lemma is to show that each term contributing to $F$ in (\ref{1a6}), 
namely $D_4, C_{2,1,1}, E_4$, and $E_2^2$,  admits  by itself an expansion of the form (\ref{1a8}). 
To show this, and to compute the contribution to $\cF_k$ from each term, we will perform the summation 
over the integers $n_r$ (but not the integers $m_r$), using the fundamental formula,
\bea
\label{3a1}
\sum _{n \in \ZZ} \, { 1 \over z+n} = - i \pi \, { 1+ e^{2 \pi i z}  \over 1- e^{2\pi i z} } 
\eea
This formula, and other formulas involving higher powers of $1/(z+n)$ that are derived from it by taking successive derivatives in $z$, are discussed Appendix \ref{appA}.    Inspection of the results 
of these summations will prove part 1 of the Lemma, and will provide explicit formulas 
for the coefficients $\cF_k$. The proof of part 2 of the Lemma is an easy application of the 
explicit formulas for $\cF_k$ derived in part 1.

\sm

The strategy for proving part 3 of the Lemma is to use extensive algebraic simplifications and rearrangements of 
the coefficients $\cF_k$ to prove that all non-harmonic contributions cancel. These calculations are considerably facilitated  by the use of MAPLE. 

\sm

To avoid a proliferation of factors of $\pi$ and $\tau_2$ while carrying out the calculation of $\cF_k$ 
described above, it will be convenient to extract from $F$ a common factor of $\tau_2^4/\pi^4$, by defining the 
following reduced functions, 
\bea
\label{3a2}
F = { \tau_2^4 \over \pi^4} \, \cF 
\hskip 0.7in 
D_4 = { \tau_2^4 \over \pi ^4} \, \cD
\hskip 0.7in 
C_{2,1,1} = {\tau_2^4 \over \pi ^4} \, \cC
\hskip 0.7in 
E_s =  {\tau_2 ^s \over \pi^s} \, \cE_s 
\eea
in terms of which the decomposition of  the Lemma takes the form, 
\bea
\label{3a3}
\cF(\tau, \bar \tau)= \cD(\tau, \bar \tau)  - 24 \, \cC(\tau, \bar \tau) - 3 \, \cE_2(\tau, \bar \tau)^2 
+ 18 \, \cE_4 (\tau, \bar \tau)
= \sum _{k=0} ^7 {\pi ^8 \over (\pi \tau_2)^k} \,  \cF_k (q, \bar q)
\eea
where $q=e^{2 \pi i \tau}$ and $\cF_k(q, \bar q)$ is an entire function of $q$ and $\bar q$.

\subsection{Expansion of $\cE_2$ and $\cE_4$}

We will now write the expression for the Eisenstein series in a form that has the structure of the power series in $\tau_2$ on the right-hand side of (\ref{3a3}), where the coefficients are functions of $q$ and $\bar q$.  We will later derive expressions  for $\cD$ and $\cC$ in the same format.  This representation of $\cE_s$ can be deduced from the defining relation for $E_s$  in (\ref{1a1}). 
To carry out the sum over $n$, it will be convenient to split the sum over $(m,n) \not=(0,0)$ into a contribution from $m=0$, in which case the sum over $n$ must exclude $n=0$, and the contribution from $m\not=0$, in which case the sum over $n$ runs over all integers. One obtains the following decomposition,
\bea
\label{3b1}
\cE_s (\tau, \bar \tau) = \sum _{n \not= 0} \, { 1 \over |n|^{2s} } 
+ \sum _{m \not= 0} \, \sum _{n \in \ZZ} \, { 1 \over |m\tau+n|^{2s}}
\eea
The first term on the right side of (\ref{3b1}) equals $2 \zeta (2s)$, and provides the leading behavior near the cusp $\tau_2 \to \infty$ for $\Re (s) > 1$. We shall evaluate the second term on the right side  for the special values $s=2,4$ needed here.

\sm

To compute the sum over $n$ for $\cE_2$, we decompose the argument of the sum, which is a rational function in $n$,  into partial fractions and express the result using the notation $z=m\tau$ for $m \not=0$, with $z=x+iy$ for $x,y \in \RR$, 
\bea
\label{3b2}
{ 1 \over |z +n|^4 } = - { 1 \over 4y^2 \, (z+n)^2} - { 1 \over 4y^2 \, (\bar z+n)^2} + { i \over 4 y^3 \, (z + n)} 
- { i \over 4 y^3 \, (\bar z + n)} 
\eea
The summations over $n$ may now be performed by using (\ref{3a1}) and the first formula in (\ref{A2}).
The result is,
\bea
\label{3b3}
\cE_2= { \pi^4 \over 45} + \sum _{m \not = 0} \left (
{ \pi \over 4 m^3  \tau_2^3} \, { 1+q^m \over 1- q^m} 
+ { \pi^2 \over m^2  \tau_2^2} \, { q^m \over (1-q^m)^2} + \hbox{c.c.} \right )   
\eea
Here and below, the notation c.c. denotes the addition of the complex conjugate of the entire preceding expression.
The first term on the right-hand side of (\ref{3b3}) accounts for the contribution from $m=0$, which has been evaluated, noting that $2\zeta (4) =\pi^4/45$.  

\sm

The calculation for $\cE_4$ proceeds by decomposition into partial fractions as well, 
and requires the use of (\ref{A1}), as well as all three formulas in (\ref{A2}). The result is as follows,   
\bea
\label{3b4}
\cE_4 & = & { \pi^8  \over 4725} 
+ \sum _{m \not= 0} 
\left ( { 5 \pi \over 32  m^7  \tau_2^7} \, { 1+q^m \over 1-q^m} 
+ { 5 \pi^2 \over 8 m^6  \tau_2^6 } \, { q^m \over (1-q^m)^2}  \right .
\no \\ && \hskip 0.8in \left . 
+{ \pi^3 \over 2 m^5  \tau_2^5} \, { q^m(1+q^m) \over (1-q^m)^3} 
+ { \pi^4 \over 6 m^4  \tau_2^4} \, { q^m+4q^{2m} + q^{3m} \over (1-q^m)^4} + \hbox{c.c.} \right )   
\eea
The first term is the  $m=0$ contribution, which is equal to  $2 \zeta (8) = \pi^8/4725$. 

\sm

The expressions (\ref{3b3}) and (\ref{3b4})  for $\cE_2$ and $\cE_4$ manifestly have the structure of the representation of $\cF_k$ in terms of a  power series in $1/\tau_2$ on the right-hand side of  (\ref{3a3}). In fact, 
the contributions to $\cF_k$ from $\cE_4$ are manifestly harmonic, and thus satisfy the property of $\cF_k$
in part 3 of the Lemma.

\subsection{Expansion of $\cC$}

Obtaining an expansion of $\cC$ in a form that is analogous to the expansions of  $\cE_2, \cE_4$  is more involved. 
We start with the expression for $\cC$ as a multiple sum, which may be obtained from (\ref{1a3}), 
\bea
\label{3c1}
\cC(\tau, \bar \tau )= \sum _{(m_r,n_r) \not= (0,0)} 
{ \delta _{m,0} \delta _{n,0}  \over  |m_1  \tau  + n_1|^4 \,  |m_2  \tau  + n_2|^2 \, 
|m_3  \tau  + n_3|^2 } 
\eea
To carry out the summations over the integers $n_r$, we must again partition the contributions according to the vanishing pattern of the integers $m_r$. Note that the summand is symmetric under the permutation of the indices $2,3$, but there is no such symmetry with index 1.  As a result, $\cC$ may be decomposed as follows,
\bea
\label{3c2}
\cC (\tau, \bar \tau) =  \cC ^{(3)}  
+ \sum _{m_1 \not= 0} \left ( 2 \, \cC ^{(2)} (m_1 \tau )  
+ \cC ^{(1)} (m_1 \tau ) \right )  
+ \sum _{m_r \not= 0} \, \delta _{m,0} \, \cC ^{(0)} (m_1 \tau , m_2 \tau ,m_3 \tau ) 
\eea
where the combinatorial coefficients take into account the symmetries of the various multiple sums. 
The partial contributions are given by, 
\bea
\label{3c3}
\cC ^{(3)}  & = & 
\sum _{n_r \not= 0}  { \delta _{n,0}  \over  n_1^4 \, n_2^2 \, n_3^2 } 
\no \\
\cC ^{(2)} (z_1) & = & 
\sum _{n_2 \not= 0} \, \sum _{n_1, n_3}  { \delta _{n,0}  \over n_2^2  \,   |z_1 + n_1|^4  \,   |- z_1 + n_3|^2} 
\no \\
\cC ^{(1)} (z_1) & = & 
\sum _{n_2 \not= 0} \, \sum _{n_1, n_3}  { \delta _{n,0}  \over n_2^4 \,   |z_1 + n_1|^2 \,  |-z_1+ n_3|^2 } 
\no \\
\cC ^{(0)} (z_1, z_2,z_3) & = & 
\sum _{n_r }  { \delta _{n,0}  \over  |z_1 + n_1|^4 \,  |z_2 + n_2|^2 \,   |z_3 + n_3|^2 } 
\eea
Conservation of $m_r$ imposes the condition $z_1+z_2+z_3=0$ with $z_r \not= 0$ in  $\cC ^{(0)}$. 
We have retained the dependence on all three variables $z_r$, as the associated manifest permutation 
symmetry in their indices will be convenient for later purposes.

\sm

The calculation of the constant  $\cC^{(3)}$  is presented in Appendix \ref{appA}, and we find,
\bea
\label{3c4}
\cC^{(3)} = { 2 \pi^8 \over 14175}
\eea
The evaluation of the remaining $\cC$-functions proceeds in analogy with the evaluation of $\cE_2$ 
given earlier, and has been performed using MAPLE.  In expressing the results,  we shall use the 
notation $q_r=e^{2\pi i z_r}$  and $y_r=\Im(z_r)$.  The results are as follows,
\bea
\label{3c5}
\cC ^{(2)}(z_1)
& = & 
\left ( { \pi^2 \over 16 y_1^6} + { \pi^4 \over 12 y_1^4} \right ) 
\left ( { q_1 \over (1-q_1)^2} + { \bar q _1 \over (1-\bar q_1)^2} \right )
- { \pi^2 \over 4 y_1^6} { 1 + q_1 \bar q_1  \over |1-q_1|^2}
\no \\ && + 
\left ( { 5 \pi \over 32 y_1^7} + { \pi^3 \over 8 y_1^5} \right )   { 1-q_1\bar q_1 \over |1-q_1|^2} 
-{ \pi^3 \over 8 y_1^5} { (q_1+\bar q_1)(1-q_1\bar q_1) \over |1-q_1|^4 } 
\eea
and
\bea
\label{3c6}
\cC ^{(1)}(z_1) 
= 
\left ({ \pi^5 \over 90 \, y_1^3} - { \pi^3 \over 24 \, y_1^5}  - {\pi \over 32 y_1^7} \right )
{ 1 - q_1 \bar q_1 \over |1-q_1|^2}
+ { \pi^2 \over 16 \, y_1^6} \, { 1 + q_1 \bar q_1 \over |1-q_1|^2}   
\eea
Finally, the most involved part is $\cC^{(0)}$, which is given as follows,
 \bea
 \label{3c7}
\cC ^{(0)} (z_1, z_2, z_3) & = & 
 {\pi^4 \over 4 \, y_1^2 \, y_2 \, y_3} { 1+q_1\over (1-q_1)^2 (1-q_2) (1 -q_3)} 
\\ &&
+ {\pi^3 \over 8 \, y_1^3 \, y_2^2} { q_1 (1+\bar q_2) \over (1-q_1)^2(1-\bar q_2)} 
+ { \pi^3 (3y_1+2y_2) \over 8 \, y_1^3 \, y_2 \, y_3^2} { q_1(1+q_2) \over (1-q_1)^2 (1-q_2)}
\no \\ &&
- { \pi^2 (6 y_1^2 + 8 y_1 y_2 + 3 y_2^2) \over 64 \, y_1^4 \, y_2 \, y_3^3} 
{ (1 +q_1) (1+q_2) \over (1-q_1) (1-q_2)} 
-{\pi^2 (y_1-3y_2) \over 64 \, y_1^4 \, y_2^3} { (1+q_1)(1+\bar q_2) \over (1-q_1) (1-\bar q_2)}
\no \\ &&
+{ \pi^2 \over 64 \, y_2^3 \, y_3^3} { q_2 + \bar q_3 \over (1-q_2) (1-\bar q_3)} 
+ (2\leftrightarrow 3) +  \hbox{c.c.}
\no
\eea
The notation $(2\leftrightarrow 3)$ denotes the addition of the  contribution obtained by interchanging indices 2 and 3
of the entire expression, while c.c.  denotes the addition of the complex conjugate of the entire expression.
After substituting $z _r = m_r \tau$ and $ q_r = q^{m_r}$ for $r=1,2,3$ into $\cC^{(2)}$, 
$\cC^{(1)}$, and $\cC^{(0)}$, we see that the contribution of $\cC$ to $\cF$  has the same form as the expansion in powers of $1/\tau_2$ on the right-hand side of (\ref{3a3}), thereby proving part 1 of the Lemma for $\cC$.

\subsection{Expanding $\cD$}

The expansion of $\cD$ proceeds in analogy with the expansions used above for $\cE_2, \cE_4$, and $\cC$. We start with the expression for $\cD$ as a multiple sum, which may be derived from (\ref{1a5}),
\bea
\label{3d1}
\cD( \tau, \bar \tau )=\sum _{(m_r, n_r) \not = (0,0)} \delta _{m,0} \, \delta _{n,0} \,
\prod _{r=1}^4 { 1 \over | m_r  \tau  + n_r|^2}   
\eea
where $m=m_1+m_2+m_3+m_4$ and $n=n_1+n_2+n_3+n_4$. To carry out the summations over the integers $n_r$, we partition the contributions according to the vanishing pattern of the integers $m_r$.  As a result, $\cD$ may be decomposed as follows,
\bea
\label{3d2}
\cD (\tau, \bar \tau) & = & 
\cD ^{(4)} +  \sum _{m_1 \not= 0} 6 \, \cD ^{(2)} (m_1  \tau) 
+  \sum _{m_2, m_3, m_4 \not= 0} 4 \, \delta _{m,0} \, \cD ^{(1)} (m_2  \tau  ,m_3  \tau  ,m_4  \tau ) 
\no \\ &&
+ \sum _{m_1, m_2, m_3, m_4 \not=0} \delta _{m,0} \, \cD^{(0)} (m_1  \tau, m_2  \tau, m_3  \tau, m_4  \tau ) 
\eea
By a slight abuse of notation, $m$ will stand for the sum of all summation variables $m_r$, whether that number of variables is three as in the third term on the right side, or four as in the fourth term.
The reduced contributions are given by,
\bea
\label{3d3}
\cD ^{(4)}  
& = &
\sum _{n_r \not= 0}   {\delta _{n,0}  \over n_1^2 \, n_2^2 \, n_3^2 \, n_4^2}    
\no \\
\cD ^{(2)} (z_1) 
& = &
\sum _{n_3, n_4 \not= 0}  \, \sum _{n_1, n_2 \in \ZZ}  {\delta _{n,0}  \over n_3^2 \, n_4^2 \, 
  | z_1  + n_1| ^2 \,  | - z_1 + n_2|^2}
\no \\
\cD ^{(1)} (z_2,z_3,z_4) 
& = &
\sum _{n_1 \not= 0}  \, \sum _{n_2,n_3, n_4 \in \ZZ}  {\delta _{n,0}  \over n_1^2} \,
 \prod _{r=2}^4 { 1 \over  ( z_r  + n_r)(\bar z_r + n_r)}
 \no \\
\cD^{(0)} (z_1,z_2,z_3,z_4) 
& = &
\sum _{n_r \in \ZZ}  \delta _{n,0} \,
\prod _{r=1}^4 { 1 \over  ( z_r  + n_r)(\bar z_r + n_r)}
\eea
In each case we impose the $m_r$ conservation condition, $\sum_r z_r=0$.  

\sm

$\bullet$ The contribution from the constant $\cD^{(4)}$ is evaluated in Appendix \ref{appA}, and we find, 
\bea
\label{3d4}
\cD^{(4)} = { \pi^8 \over 945}
\eea
There is no $\cD^{(3)}$ contribution because imposing the vanishing of three of the integers $m_r$ implies the vanishing of the fourth one as well. The remaining $\cD$-functions are as follows.

\sm

$\bullet$ The contributions to $\cD^{(2)}$ partitions into two parts: one from $n_1+n_2=-n_3-n_4 \not=0$, 
and the other from $n_1+n_2=-n_3-n_4 =0$. Collecting both gives, 
\bea
\label{3d5}
\cD^{(2)} (z_1) & = & 
\left ( {  \pi^5 \over 18 y_1^3} + { \pi^3 \over 3 y_1^5} +{ 3 \pi \over 16 y_1^7} \right )  
 { 1-q_1\bar q_1 \over |1-q_1|^2}
\\ && 
- \left (  { \pi^4 \over 6 y_1^4}  + { 3 \pi^2 \over 8 y_1^6} \right ) 
{ 1+q_1\bar q_1 \over |1-q_1|^2}
+{ \pi^6 \over 45 y_1^2} \left ( { q_1 \over (1-q_1)^2} + { \bar q_1 \over (1-\bar q_1)^2} \right )   
\no
\eea 

\sm

$\bullet$ The contribution  $\cD^{(1)} (z_1, z_3, z_4)$ may be expressed as follows,
\bea
\label{3d6}
\cD^{(1)} (z_2,z_3,z_4) & = & \Phi (z_2,z_3)+\Phi (z_3,z_4)+\Phi (z_4,z_2)    
\eea
where we have defined the function, 
\bea
\Phi (z_r,z_s) & = &  { \pi ^4 \over 12 y_r^2 \, y_s^2} \, \Re \left ( { 1+q_r \over  1-q_r}  \right ) 
\Re \left ( { 1+q_s \over  1-q_s}  \right ) 
\no \\ &&
- { \pi^3 \over 4 y_r \, y_s \, y_t^3} \, \Re \left ( { q_r q_s + \bar q_t \over (1-q_r)(1-q_s)(1-\bar q_t) } \right )
\no \\ &&
+{ \pi^2 \over 8 y_r y_s y_t^4} \, \Re \left ( { q_r q_s - \bar q_t \over (1-q_r)(1-q_s)(1-\bar q_t)} \right )
\eea
with $t \in \{2,3,4\}$ and  $t \notin \{ r,s \} $, so that 
$y_r+y_s+y_t=0$ and $q_r q_s q_t =1$.

\sm

$\bullet$ The contribution from $\cD^{(0)}$ must be partitioned into the part
for which all pairs satisfy $z_r+z_s \not= 0$ for $r \not= s$, and the part
for which we have $z_1+z_2=z_3+z_4=0$ or permutations thereof.  The results may be collected as follows.
\bea
\label{3d8}
\cD^{(0)} (z_1, z_2, z_3, z_4) & = & 
{\pi ^4 \over y_1y_2y_3y_4} \, { 1 \over (1-q_1)(1-q_2)(1-q_3)(1-q_4)} 
\no \\ &&
- { \pi^3 (1-\delta _{y_1+y_2,0}) \over 4 y_1y_2y_3y_4 (y_1+y_2)} \, 
{ (q_1 q_2-\bar q_3 \bar q_4) \over (1-q_1) (1-q_2) (1-\bar q_3) (1-\bar q_4)} 
+\hbox{ 2 perms}
\no \\ &&
+ { \pi^4 \delta _{y_1+y_2,0} \over y_1^2 y_3^2 } { q_1 \bar q_3 \over (1-q_1) ^2 (1-\bar q_3)^2} +\hbox{ 2 perms}
\no \\ &&
- { \pi^3 \over 4 y_1 ^2 y_2y_3y_4 } \, 
{ (q_1- \bar q_2 \bar q_3 \bar q_4 ) \over (1-q_1) (1-\bar q_2) (1-\bar q_3) (1-\bar q_4)} 
+ \hbox{3 perms} 
\no \\ && \no \\ &&
+ \hbox{ complex conjugate}
\eea
The terms listed on the second and third lines correspond to the partition $(12|34)$; the two  permutations to be added correspond to the partitions $(13|24)$ and $(14|23)$. The term listed on the fourth line corresponds to the partition $(1|234)$; the three permutations to be added correspond to the  partitions $(2|134)$, $(3|124)$, and $(4|123)$. The complex conjugate, which interchanges $q_r$ with $\bar q_r$ and leaves $y_r$ invariant, of the entire expression is to be added.

\subsection{Summary of contributions to $\cF_k$}

A summary of the contributions of the various non-vanishing components of $\cD$, $\cC$, $\cE_2^2$ and $\cE_4$ to $\cF$, is presented in table 1.
 
\begin{table}[htdp]
\begin{center}
\begin{tabular}{|c||c |  c | c | c  | c | c | c   |} \hline
$\cD^{(0)}$    &    &  &  & $\tau_2^{-4}$ & $\tau_2^{-5}$ &  &   \\ \hline
$\cD^{(1)}$   &    &  &  & $\tau_2^{-4}$ & $\tau_2^{-5}$ & $\tau_2^{-6}$ &   \\ \hline
$\cD^{(2)}$   &    & $\tau_2^{-2}$ & $\tau_2^{-3}$ & $\tau_2^{-4}$ & $\tau_2^{-5}$ & $\tau_2^{-6}$ & $\tau_2^{-7}$  \\ \hline
$\cD^{(4)}$   &  $\tau_2^{0}$   &  &  &  &  &  &   \\ \hline
$\cC^{(0)}$      &    &  &  & $\tau_2^{-4}$  & $\tau_2^{-5}$ & $\tau_2^{-6}$ &   \\ \hline
$\cC^{(1)}$      &    &  & $\tau_2^{-3}$ &  & $\tau_2^{-5}$ & $\tau_2^{-6}$ & $\tau_2^{-7}$  \\ \hline
$\cC^{(2)}$      &    &  &  &  & $\tau_2^{-5}$ & $\tau_2^{-6}$ & $\tau_2^{-7}$  \\ \hline
$\cC^{(3)}$      &  $\tau_2^{0}$   &  &  &  &  &  &   \\ \hline
$\cE_2^2$       &  $\tau_2^{0}$   & $\tau_2^{-2}$ & $\tau_2^{-3}$ & $\tau_2^{-4}$ & $\tau_2^{-5}$ & $\tau_2^{-6}$  &   \\ \hline
$\cE_4$           &  $\tau_2^{0}$   &  &  & $\tau_2^{-4}$ & $\tau_2^{-5}$ & $\tau_2^{-6}$ & $\tau_2^{-7}$  \\ \hline \hline
\end{tabular}
\end{center}
\label{table1}
\caption{The non-vanishing powers of $\tau_2$ in the expansion of the contributions to $\cF$.}
\end{table}
\sm

Substituting  $z_r = m_r \tau$ and $q_r = q^{m_r}$ into the expressions (\ref{3d5}), (\ref{3d6}) and (\ref{3d8}), results in  expansions of the functions $\cD^{(2)}$,  $\cD^{(1)}$, and $\cD^{(0)}$ in the form of power series in $1/\tau_2$.  We therefore see that the contribution of  $\cD$ to  $\cF$ is again expressed as a power series in $1/\tau_2$ of the same form as the right-hand side of (\ref{3a3}).   
 This completes the proof of part 1  of the Lemma, since now all 
contributions, namely $\cD$, $\cC$, $\cE_2^2$ and $\cE_4$ have been proven to have the 
form given in (\ref{3a3}).

\subsection{Vanishing of the contributions $\cF_k$ with $k=0,1,2,3,7$}
\label{sec32}

By inspection of the above results, one readily shows the vanishing of the coefficients $\cF_k$
for $k=0,1,2,3,7$. The arguments are as follows.
\begin{itemize}
\itemsep=0.0 in
\item  The cancellation of $\cF_0$ follows by checking the pure power terms, as was done in \cite{D'Hoker:2015foa},  
since none of these terms depend on $q$ or $\bar q$. The result of  \cite{D'Hoker:2015foa} is double checked by adding the contributions  computed to this order, which come from the terms listed in the first column of table 1,
\bea
\label{3f1}
\cD^{(4)} - 24 \cC^{(3)} - 3 \times 4 \times \zeta(4)^2 + 18 \times 2 \times \zeta (8)=0
\eea
\item  The cancellation of $\cF_1$ results from the observation that no terms of order $\tau_2^{-1}$
arise  in any of the contributions to $\cF$.
\item The cancellation of $\cF_2$ results from the combination of just two contributions, namely $\cD^{(2)}$ and $\cE_2^2$. They both have the same functional dependence on $q$ and $\bar q$,
\bea
\label{3f2}
\sum _{m \not= 0} { 1 \over m^2} \, \left (  { q^m \over (1-q^m)^2} + { \bar q^m \over (1- \bar q^m)^2} \right )
\eea
Upon properly including the combinatorial factors, we find that their sum cancels.  
\item The cancellation of $\cF_3$ results from combining the three terms of  order $\tau_2^{-3}$ in $\cD^{(2)}$, $\cC^{(1)}$, and $\cE_2^2$. They all have the same functional dependence on $q, \bar q$, given by,  
\bea
\label{3f3}
\sum _{m \not= 0}  { 1 \over m^3} \, { 1 - q^m \bar q^m \over (1-q^m)(1-\bar q^m)}    
\eea
The coefficients are as follows (including combinatorial factors),
\bea
\label{3f4}
6 \times { 1 \over 18} - 24 \times {1 \over 90} - 3 \times 2 \times {1 \over 45} \times {1 \over 2} =0
\eea
\item  Finally, the cancellation of $\cF_7$ results from combining the four terms of  order $\tau_2^{-7}$,
namely  in $\cD^{(2)}$, $\cC^{(1)}$, $\cC^{(2)}$, and $\cE_4$. These four contributions  all have the same functional dependence on $q, \bar q$, given by,
\bea
\label{3f5}
\sum _{m \not= 0} { 1 \over m^7} \, { 1 - q^m \bar q^m \over (1-q^m)(1-\bar q^m)} 
\eea
The coefficients are as follows (including combinatorial factors),
\bea
\label{3f6}
6 \times {3 \over 16} - 24 \times \left ( - { 1 \over 32} \right ) - 24 \times  2 \times  { 5  \over 32} 
+ 18 \times  { 5 \over 16} =0
\eea
\end{itemize}
This cancellation completes the proof of part 2 of the Lemma that $\cF_k =0$ for $k=0,1,2,3,7$.

\subsection{Harmonic structure of $\cF_k$ for $k=4,5,6$}

The analysis of the remaining terms in $\cF$, namely, $\cF_4$, $\cF_5$ and  $\cF_6$ is considerably 
more complicated, and we relegate the detailed discussion to appendix~\ref{appB}.  
 
 \sm
 
There, we will  begin by collecting all the terms that contribute to $\cF_k$ for $k=4,5,6$.   
Some contributions to $\cF_k$ are {\sl manifestly harmonic} in view of the fact that they enter as the sum
of a holomorphic function of $q$ and its complex conjugate. Other contributions are not 
manifestly harmonic, and will be collected into a sub-contribution denoted by $\cF_k ^{{\rm nh}}$.  
Using extensive algebraic rearrangements of the sum of the terms in  $\cF_k^{{\rm nh}}$ it will be 
shown that all non-harmonic dependence in $\cF_k^{{\rm nh}}$ in fact cancels, so that  $\cF_k^{{\rm nh}}$ also contributes harmonic terms, thereby proving that $\cF_k$ is a purely real harmonic expression of the form,
\bea
\label{3g1}
\cF_k(\tau,\bar\tau) = \f_k (q)+ \f_k(\bar q)
\label{harmsum}
\eea
The explicit form of $\f_k(q)$ will be presented in Appendix \ref{appB}. By inspection, we will show that
the functions $\f_k (q)$ satisfy the conjugation properties of (\ref{1a10}) of the Lemma given in the introduction. 
From the first equation of (\ref{1a10}), we conclude that $\f_k$ is a real function of $q$, 
whose Taylor series expansion in powers of $q$ has real coefficients. 
The combination of these two properties implies that, when viewed as a function of $\tau$, the function 
$\psi _k (\tau)=\f_k(q)$ has the following equivalent conjugation properties,
\bea
\label{3g3}
\psi _k (\tau) = \f _k (e^{2 \pi i \tau}) & \hskip 0.7in & \overline{ \psi _k (\tau)} ~ ~ = (-)^k \, \psi _k (\bar \tau)
\no \\ &&
\psi _k (- \tau) = (-)^k \, \psi _k (\tau)
\eea
Using these explicit formulas, and MAPLE based calculations, we have shown that $\f_k(q) = \cO(q^{300})$. In the subsequent section, we shall produce a proof of the vanishing of $\f_k$ to all orders in $q$ by exploiting the modular invariance of $F(\tau, \bar \tau)$.

\section{Proof of the Theorem}
\setcounter{equation}{0}
\label{sec4}

The  Lemma, presented in the introduction and proven in the preceding section, 
strongly  constrains the structure of the non-holomorphic function $F(\tau, \bar \tau)$, 
introduced in (\ref{1a6}). To prove the Theorem presented in (\ref{1a7})  
we need to prove that $F=0$. Our strategy will be to combine the strongly constrained
structure of $F$, derived from explicit computations of $\cF_k$ in the preceding section, with 
the modular invariance property of $F$. An alternative derivation due to Deligne will be presented in Appendix~D. \footnote{We are grateful to Pierre Deligne for communicating his derivation to us following the appearance of the first version of this paper on the archive.}

\subsection{Modular properties of  $\psi _k (\tau)$}

We begin by  describing the strong structural constraint derived in the Lemma in a manner that 
will be suitable for investigating the modular properties of $\psi _k (\tau) = \f_k(q)$. To this end, 
we restore the normalizations of (\ref{3a2}), and use the decomposition provided by the 
Lemma in (\ref{3a3}) and (\ref{harmsum}) to arrive at the  following expression, 
\bea
\label{4a1}
\pi^4 F(\tau, \bar \tau) = H(\tau, y) + H(\bar \tau, y)
\eea
We use the notation $y = \tau_2$ for later convenience.  The function $H$ may be expressed in
terms of the  functions $\psi _k(\tau)$ for $k=4,5,6$, which are holomorphic in $\tau \in \mH$,  by the relation,
\bea
\label{4a2}
H(\tau, y)  =  \psi_4(\tau) + { 1 \over y} \psi_5 (\tau) + { 1 \over y^2} \psi_6 (\tau)
\eea
It will generally be more convenient to express the modular properties on the form of $H$ in terms of $\psi _k$, 
since then all dependence will be manifestly in terms of the modulus $\tau$. The conjugation properties of 
(\ref{1a10}) for $\f_k(q)$ and (\ref{3g3}) for  $\psi_k (\tau) $ guarantee that $H$ satisfies the following 
conjugation properties, 
\bea
\label{4a2a}
\overline{H(\tau, y)} & = & H(\bar \tau, y)
\no \\
H(-\tau, -y) & = & H(\tau, y)
\eea
We shall sometimes refer to the functions $H$ as {\sl almost holomorphic}.

\sm

Next, we analyze the constraints on $\f_k(q)=\psi _k(\tau)$ imposed by the modular invariance of $F$. Since $\f_k(q)$ 
admits a $q$-expansion near the cusp, it is an entire function of $q$ and hence invariant under the modular transformation $T: \tau \to \tau+1$, which implies the periodicity,
\bea
\psi_k(\tau + 1) & = & \psi_k(\tau)
\no \\
H(\tau+1, y) & = & H(\tau, y)
\eea
Modular invariance of $F$ under the transformation $S: \tau \to - 1/\tau$ requires, 
\bea
F(\tau,\bar\tau)= F(-1/\tau,-1/\bar\tau)
\label{4a3}
\eea
Making use of the identities 
  \bea
{|\tau|^2\over \tau_2} = { \tau^2 \over \tau_2} - 2 i \tau = { \bar \tau^2 \over \tau_2} + 2 i \bar \tau
\label{4a4}
\eea
allows us to express $ F(-1/\tau,-1/\bar \tau)$ as the sum of an {\sl almost holomorphic function} $H^S(\tau,y)$ 
and its complex conjugate $H^S(\bar \tau,y)$,
\bea
\pi^4 F(-1/\tau,-1/\bar \tau) = H^S(\tau,y) + H^S(\bar\tau,y)
 \label{4a5}
 \eea
where $H^S$ may be chosen to have the same functional form as $H$ in (\ref{4a2}) (the decomposition is not unique, as may be seen by adding to $H^S$ an imaginary function of $y$ only), 
 \bea
 \label{4a6}
 H^S(\tau,y) &=&  \psi_4(-1/\tau)-2 i \tau \psi_5(-1/\tau) - 4 \tau^2 \psi_6(-1/\tau)\no\\
 &&+ {1\over y} \Big ( \tau^2\psi_5(-1/\tau)  -4i \tau^3 \psi_6(-1/\tau) \Big )  +{1\over y^2} \tau^4 \psi_6(-1/\tau)
\eea
It is clear by inspection that $H^S$ satisfies the same conjugation relations as $H$ does in (\ref{4a2a}).
By eliminating $F$ between (\ref{4a1}), (\ref{4a3}), and (\ref{4a5}), we obtain a relation which 
expresses the modular  $S$-invariance of $F$ in terms of the functions $H$ and $H^S$, 
\bea
\label{4a6a}
H(\tau,y) + H(\bar \tau, y) = H^S (\tau, y) + H^S (\bar \tau, y)
\eea 
An alternative representation of the same formula is given by 
\bea
\label{4a6b}
K(\tau,y) = - K(\bar\tau,y)
\eea 
where we have defined the function $K$ by, 
\bea
K(\tau,y) = H(\tau,y) - H^S(\tau,y)
\label{4a7}
\eea
In view of relation (\ref{4a6b}), $K$ is purely imaginary for all $\tau$ and $y$, while in view of the conjugation relations of (\ref{4a2a}) for $H$ and  $H^S$,  the following conjugation relations  hold,
\bea
\label{4a7a}
\overline{K(\tau, y)} & = & K(\bar \tau, y)
\no \\
K(-\tau, -y) & = & K(\tau, y)
\eea
Furthermore, from the expressions for $H$ and $H^S$ in terms of  $\psi_k (\tau)$ and $\psi_k (-1/\tau)$, 
we see that $K(\tau, y)$ has an expansion in powers of $y$ just as $H$ and $H^S$ do,
\bea
\label{4a8}
K(\tau, y) = K_4 (\tau) + { 1 \over y} K_5(\tau) + { 1 \over y^2} K_6(\tau)
\eea
where $K_k(\tau)$ are holomorphic functions of $\tau$. 

\sm

In appendix~\ref{appC}, we show that the general form of $K$  in (\ref{4a8}), subject to the 
condition that it be purely imaginary, and satisfy the conjugation properties of (\ref{4a7a}), implies\footnote{We are very grateful to Stephen Miller for  suggesting this procedure for constraining $K$, and to Pierre Deligne for pointing out an incompleteness in the proof of $K=0$ given in the first version of this paper.} 
that  $K$ is restricted to the following form, 
\bea
K(\tau,y) = A + 2i B\tau + \frac{1}{y}(C+ iA\tau - 3B \tau^2) + \frac{1}{y^2} (i C \tau - i B \tau^3)
\label{Kres}
\eea
for some real constants $A, B, C$. This form for $K$ extends to the lower half plane with the signs
of $B$ and $C$ are reversed, and $A$ unchanged, upon using the second equation in (\ref{4a7a}).

\sm

The relation between $H$, $H^S$ and $K$ depends on $y$ and $\tau$.  These may be viewed as independent variables, as follows from writing $y^2 K(\tau,y)$ as a power series in $\tau$ and $\bar \tau$.  Consequently, the various powers in $y$ may be identified by comparing (\ref{4a7}) and (\ref{Kres})\footnote{The combinations of $\psi_k(\tau)$ arising in (\ref{4a10}) are closely connected to period polynomials \cite{KZ,ZP}.},
\bea
\psi_4(\tau)- \psi_4(-1/\tau) + 2 i \tau \psi_5(-1/\tau) + 4 \tau^2 \psi_6(-1/\tau) &=& A + 2i B \tau 
\no \\
\psi_5(\tau) -\tau^2 \psi_5(-1/\tau)  +4i \tau^3 \psi_6(-1/\tau)&=& C+ iA\tau - 3 B \tau^2
\no \\
\psi_6(\tau) - \tau^4 \psi_6(-1/\tau) &=& i C \tau - i B \tau^3
\label{4a10}
\eea
These equations must hold for all $\tau$ in the Poincar\'e upper half space $\mH$.

\subsection{Proving the vanishing of $A,B,C$, and $\psi _4, \psi_5, \psi _6$}

We will now argue that the modular properties of $\psi_k$ are  determined by further consistency conditions, which follow from the periodicity of $\psi_k(\tau)$ under $\tau\to \tau+1$, and which restrict the coefficients $A$, $B$, $C$ on the right-hand side of (\ref{4a10}).  

\sm

Transformation of the third equation in (\ref{4a10}) under $S: \tau\to -1/\tau$, 
and subsequent multiplication by $-\tau^4$  is a symmetry of the left-hand side of the equation and interchanges $B$ and $-C$ on the right-hand side.  We therefore see that consistency requires $C=-B$.

\sm

Next we set $C=-B$ and consider the transformation of the second and third equations in 
(\ref{4a10}) under $ST^{-1}: \tau \to -1/(\tau-1)$, and its square $(ST^{-1})^2$ (note that the square 
of $ST^{-1}$ equals the inverse of $ST^{-1}$ by the property $(ST^{-1})^3=I$).  
The transformations of the last equation in (\ref{4a10})  under $ST^{-1}$ and its square are given respectively by,
\bea
(\tau-1)^4\, \psi_6(-1/(\tau-1)) - \psi_6(\tau) & = & i B (\tau-1)^3 + i B (\tau-1) 
\no \\
\tau^4\, \psi_6(-1/\tau) - (\tau-1)^4\, \psi_6(-1/(\tau -1)) & = & - i B \tau^3(\tau-1)  + i B \tau (\tau-1)^3
\label{conds3a}
\eea
where the invariance of $\psi_6(\tau)$ under $\tau \to \tau +1$ has been used.
Adding the sum of both equations in (\ref{conds3a}) to the last equation in (\ref{4a10}) gives the relation
$2iB( 1- \tau + \tau^2)^2 =0$,  which implies that $B=C=0$. 
As a result, the third equation in (\ref{4a10}) reduces to, 
\bea
\psi_6(\tau) = \tau^4 \psi_6(-1/\tau)
\label{phi6mod}
\eea
Since no holomorphic modular form of weight $-4$ exists, we  have $\psi_6(\tau)=0$. 

\sm

Having set $B=C=\psi_6=0$, the transformations  of the second equation 
in (\ref{4a10}) under $ST^{-1}$ and its square are given respectively by,
\bea
(\tau-1)^2\, \psi_5(-1/(\tau-1)) - \psi_5(\tau) & = &  - i A\,(\tau-1)
\no \\
\tau^2\, \psi_5(-1/\tau) -(\tau-1)^2\, \psi_5(-1/(\tau-1)) & = &  i A\,\tau (\tau-1)
\label{conds2b}
\eea
Adding the sum of both equations in (\ref{conds2b}) to the second equation of (\ref{4a10}) gives the relation
$iA( 1- \tau + \tau^2) =0$, which implies $A=0$. Hence, the second equation in (\ref{4a10}) reduces to, 
\bea
\psi_5(\tau) = \tau^2 \psi_5(-1/\tau)
\label{phi5mod}
\eea
Since no holomorphic modular form of weight $-2$ exists,  we  have $\psi_5(\tau)=0$.  
Finally we see that the first equation in (\ref{4a10})  with $A=B=C= \psi_6(\tau)=\psi_5(\tau)=0$  implies 
that $\psi_4(\tau)$ must be a holomorphic modular form of weight $0$, which must be constant.
Since we know that the constant term of $\psi _4 (\tau)$ is zero, it follows that $\psi _4 (\tau)=0$.

\sm

Since the functions $\psi_k(\tau) = \f_k(q)$ vanish for $k=4,5,6$, it follows from (\ref{harmsum}) that 
$\cF_k$ vanishes for those values of $k$. Together with the results of section \ref{sec32} for $k=0,1,2,3,7$, 
we conclude that $\cF_k$ vanishes for $k=0,1, \cdots, 7$. Therefore, in view of (\ref{1a8}), we have, finally, 
proven the Theorem (\ref{1a7}) of the introduction.

\section{Holomorphic corollaries}
\setcounter{equation}{0}
\label{sec5}

In section \ref{sec4}, we proved that the holomorphic functions $\f_4(q), \f_5(q)$, and $\f_6(q)$
are modular forms of respective weights $0, -2$, and $-4$. Therefore, using the known asymptotic behavior of $\f_4$ near the cusp, it follows that  $\f_4, \f_5$ and $\f_6$ must individually vanish.   These results were obtained by exploiting 
the holomorphicity of $\f_4,\f_5$ and $\f_6$, and  the vanishing of $\cF_k$ for $k=0,1,2,3,7$ established in 
section \ref{sec32},  along with the modular invariance of $F$. 

\sm

In addition, we obtained {\sl explicit expressions} for $\f_4, \f_5, \f_6$ respectively in (\ref{B7}), (\ref{B22}),
and (\ref{B31}), which, in view of the above result,  manifestly satisfy the corollaries:
\begin{enumerate}
\itemsep =0in
\item[$\bullet$] They are modular forms of weights $0, -2, -4$ respectively;
\item [$\bullet$] They therefore vanish as functions of $q$.
\end{enumerate}
Neither of these properties is manifest from the explicit expressions for $\f_4, \f_5, \f_6$ in (\ref{B7}), (\ref{B22}),
and (\ref{B31}), and we have not succeeded in proving either of these properties directly from the explicit 
expressions of $\f_4, \f_5, \f_6$ by analytical combinatorial methods.

\sm

Remarkably, the identity $\f_6(q)=0$ may be split up into two simpler sums, 
\bea
\label{5a1}
\f_6 (q) = \f_6 ^{(1)} (q) + \f_6 ^{(2)} (q)
\eea
where 
\bea
\label{5a2}
\f_6^{(1)} (q) = { 3 \over 4} \zeta (6) + \sum _{m_1+m_2 \not= 0}' { 9 \over 8 m_1^2 m_2^2 (m_1+m_2)^2}
\, { (1+q^{m_1} ) (1 + q^{m_2}) \over (1-q^{m_1} ) (1 - q^{m_2})} 
 \eea
 and 
\bea
\label{5a3}
\f_6^{(2)}  (q) & = &
- {15 \over 8}  \zeta (6) - \sum _{m} ' { 9 \over 4 m^6 } \, { q^m \over ( 1- q^m)^2} 
+ \sum _{m_1,m_2}' { 3  \over 16 m_1^3 m_2^3} \, 
{ (1+q^{m_1})(1+q^{m_2})  \over (1- q^{m_1})(1-q^{m_2}) } 
\no \\ && 
+  \sum _{m_1+m_2 \not= 0} ' { 3   \over 16 m_1 m_2 (m_1+m_2)^4} 
  { (1+q^{m_1}) (1 + q^{m_2}) \over (1-q^{m_1}) (1-q^{m_2}) } 
\eea
each of which vanishes separately. We have verified the validity of these identities using MAPLE 
up to order $\cO(q^{400})$. In addition, the identity $\f_6 ^{(1)}(q)=0$ may be proven by 
simple combinatorial arguments. Since the proof of the Theorem has provided a proof for $\f_6(q)=0$,
it follows that the identity $\f_6 ^{(2)}(q)=0$ is also proven.

\sm

To prove $\f_6^{(1)}(q)=0$, we evaluate the sum,
\bea
\phi_6 =  { 2 \over 3} \zeta (6) + \sum _{m_1+m_2 \not=0} ^\prime {  f(q^{m_1})  \,  f(q^{m_2})   \over m_1^2 m_2^2 (m_1+m_2)^2} \, 
\eea
where $f(x)=(1+x)/(1-x)$. Using the symmetry property $f(1/x)=-f(x)$ of the function $f(x)$,  
we recast the sum as follows,
\bea
\phi_6 = {2\over 3}\zeta(6)+ 2\sum_{m_1,m_2\geq1} \, {
 f(q^{m_1}) f(q^{m_2}) -  f(q^{m_1})  f(q^{m_1+m_2})  -   f(q^{m_2}) f(q^{m_1+m_2}) 
\over  m_1^2m_2^2(m_1+m_2)^2} 
\eea
Using the algebraic identity $f(x)f(y)-f(x)f(xy)-f(y)f(xy)=-1$, the numerator in the sum equals $-1$,
and the summand is therefore independent of $q$, 
\bea
\phi_6 =   {2\over 3}\zeta(6)- 2\sum_{m_1,m_2\geq1} \, {1\over
    m_1^2m_2^2(m_1+m_2)^2}\
\eea
Repeated use of the partial fraction identity for positive integers $a,b>0$,
\bea
  \label{e:Id3}
  {1\over m^a n^b}= \sum_{r=b}^{a+b-1} {{r-1\choose
        b-1}\over (m_1+m_2)^r m_1^{a+b-r}}+\sum_{r=a}^{a+b-1}
{{r-1\choose
        a-1}\over (m_1+m_2)^r m_2^{a+b-r}} 
\eea
leads to the following evaluation of the general sums, 
\bea
   \sum_{m_1,m_2\geq1} {1\over m_1^a
     m_2^b (m_1+m_2)^c}=
 \sum_{r+s=a+b\atop r,s>0} \,\left( {r-1\choose a-1}
    \, \zeta(c+r,s)+   {r-1\choose b-1}
    \, \zeta(c+r,s) \right)
\eea
where the multi-zeta function $\zeta (r_1, r_2)$ is defined by, 
\bea
  \zeta(r_1,r_2)=\sum_{0<n_2<n_1} {1\over n_1^{r_1} n_2^{r_2}}  
\eea
Therefore we find that, 
\bea
\phi_6=  {2\over3}  \zeta (6) - 4 \zeta (4,2) - 8 \zeta (5,1) =0
\eea

\sm

It is easy to show that there is no analogous way of partitioning the sum of the six terms in the explicit 
equation $\f_4(q)=0$ into 
separate identities, as we  did for $\f_6(q)$. To see this, one can simply truncate the $q$-expansion 
of each term to the first six non-trivial orders, and show that no linear combination other than their total 
sum can vanish.
Thus, the identity $\f_4(q)=0$ cannot be reduced further. The holds true for $\f_5(q)=0$.

\section{Discussion}
\setcounter{equation}{0}
\label{sec6}

The proof of the Theorem confirms the conjectured relationship that expresses  $D_4$ as a polynomial 
in functions of lower depth, namely, the functions $C_{2,1,1}$, $E_4$ and $E_2$.   
The conjecture made in   \cite{D'Hoker:2015foa} was based on the analysis of the first two lowest powers of 
$q= e^{2\pi i \tau}$ in the expansion of these modular functions near the cusp $\tau_2\to \infty$.  
Each of these terms was accompanied by a Laurent polynomial in  $\tau_2$  so the conjecture was based on 
matching a number of leading and sub-leading  coefficients.  Analogous conjectures were made, 
based on similar asymptotic analysis, that related each of the weight $w=5$ functions,  
$D_5$, $D_{3,1,1}$, and $D_{2,2,1}$, to polynomials in modular functions of lower depth.  
Although the proof of the $D_4$ conjecture suggests the validity of the  weight 5 conjectures  
in (\ref{D5}), the methods used in this paper may be too cumbersome to be applied systematically to these cases.
 
 \sm
 
In a separate paper we will present an alternative formulation of the Feynman graphs in which the 
modular functions are expressed in terms of single-valued multiple elliptical poly-logs based on 
generalized Bloch--Wigner polylogarithms discussed in \cite{Zagier:1990}.   It seems likely that this 
approach will lead to a more general analysis of the properties of the modular functions that arise in 
the low energy expansion and we are hopeful this will lead to an understanding of higher-weight terms 
and possibly to   the complete one-loop amplitude.  Separately, there has been some progress in 
expressing one-loop amplitudes in terms of multiple elliptical poly-logs  \cite{Broedel:2014vla} of the 
type discussed,  for example, in \cite{Brown:2011} and it would be interesting to discover the relationship 
of these to the closed string expansion under discussion in this paper.

\sm

Finally, a natural generalization of the questions addressed here and in \cite{D'Hoker:2015foa} is to the case
of genus two and higher. In fact, the study of the modular properties of the low energy expansion for the 
two-loop four-graviton superstring amplitude in \cite{D'Hoker:2013eea,D'Hoker:2014gfa} was a direct motivation for the 
investigations  in \cite{D'Hoker:2015foa}.
It would be fascinating to understand further the modular structure of two-loop partial amplitudes, 
and the possibility that modular relations, such as the ones proven here,  emerge also at two loops.

\section*{Acknowledgements}

We gratefully acknowledge correspondence with Stephen Miller, who made useful suggestions
for completing the last part of the proof of the Theorem. We are also grateful to 
Pierre Deligne for  communicating his alternative proof to the derivation of the results in section~\ref{sec4},
and for generously allowing us to include it in Appendix \ref{appD}.
The work of ED was supported in part by a grant from the National Science Foundation PHY-13-13986.
MBG acknowledges funding from the European Research Council under the European 
Community's Seventh Framework Programme (FP7/2007-2013) $\/$ ERC grant agreement no. [247252], partial support by National Science Foundation Grant No. PHYS-1066293 and the hospitality of the Aspen Center for Physics. PV acknowledges funding 
the ANR grant reference QST 12 BS05 003 01, and the CNRS grants PICS number 6430.


\appendix

\section{Calculations of the $q$ expansions}
\setcounter{equation}{0}
\label{appA}

In this section, we shall provide various details of the proof of the Lemma, part 1.

\subsection{Basic summation formulas}

In the course of the calculations we make repeated use of the summation formula,
\bea
\sum _{n\in \ZZ} { 1 \over z+n} =  - i \pi \, { 1+q \over 1-q}   
\label{A1}
\eea
where we shall use the convenient abbreviation $q=e^{2\pi i z}$. The sum is only conditionally convergent
and, following Eisenstein, should be understood as defined by the limit as $N \to \infty$ of the cutoff sum 
with $-N \leq n \leq N$.  The identity (\ref{A1}) is evident from the equality of the its residues at the poles 
located at $z= -n$ for $n\in \ZZ$, and the vanishing of both sides at $z=1/2$. Further identities that will be 
of use here follow by differentiation of (\ref{A1}) with respect to $z$, and we also have,
\bea
\sum _{n\in \ZZ} { 1 \over (z+n)^2} & = & - 4 \pi^2 \, { q \over (1-q)^2}
\no \\
\sum _{n \in \ZZ} { 1 \over (n+z)^3} & = & 4 i \pi^3  \, { q+q^2 \over (1-q)^3}
\no \\
\sum _{n \in \ZZ} { 1 \over (n+z)^4} & = & {8 \pi^4 \over 3}  \, { q+4q^2+q^3 \over (1-q)^4}
\label{A2}
\eea
and so on. The general formula may be expressed as follows, 
\bea
\label{A3}
\sum _{n \in \ZZ} { 1 \over (z+n)^{k+1}} = - i \pi \delta _{k,0}+ { (-2\pi i )^{k+1} \over \Gamma (k+1)} 
\sum _{\ell=1}^\infty \ell ^k q^\ell    
\eea
There are obvious variants of these formulas that will be needed as well when the summation 
over $n$ excludes the value $n=0$, and we have for example,
\bea
\sum _{n\in \ZZ}^\prime  { 1 \over z+n} =  - i \pi \, { 1+q \over 1-q}   - { 1 \over z} 
\label{A4}
\eea
where the prime superscript instructs us to omit the value $n=0$ from the sum. Formulas analogous to (\ref{A2}) and (\ref{A3}) may be derived again by successive differentiation.

\subsection{Calculations of the constants $\cC^{(3)}$ and $\cD^{(4)}$}

To compute $\cC^{(3)}$, defined in (\ref{3c3}), we eliminate $n_3=-n_1-n_2$, so that,
\bea
\label{A5}
\cC^{(3)} = \sum _{n_1\not=0} \sum _{n_2\not=0, -n_1} { 1 \over n_1^4 \, n_2^2 \, (n_1+n_2)^2}
\eea
The summation over $n_2$, for $n_1\not=0$, gives, 
\bea
\label{A6}
\sum _{n_2\not=0, -n_1} { 1 \over  n_2^2 \, (n_1+n_2)^2} = { 2 \pi^2 \over 3 n_1^2 } - { 6 \over n_1^4}
\eea
The remaining summation over $n_1$ gives, 
\bea
\label{A7}
\cC^{(3)} = \sum _{n_1\not=0}  \left ( { 2 \pi^2 \over 3 n_1^6 } - { 6 \over n_1^8} \right ) 
= { 4 \pi^2 \over 3} \zeta (6) - 12 \zeta (8)
\eea
which readily leads to the result of (\ref{3c4}).

\sm

To compute $\cD^{(4)}$, defined in (\ref{3d3}), we eliminate  $n_4=-n_1-n_2-n_3$, and partition the contributions according to whether $n_1+n_2=0$ or not, 
\bea
\label{A8}
\cD^{(4)} = \sum _{n_1\not= 0} \sum _{n_3\not=0} { 1 \over n_1^4 n_3^4} +
\sum _{n_1 \not=0} \sum _{n_2 \not=0, -n_1} \sum _{n_3 \not = 0, -n_1-n_2} { 1 \over n_1^2 n_2^2 n_3^2 (n_1+n_2+n_3)^2}
\eea
The first term readily evaluates to $4 \zeta (4)^2$, while the sum over $n_3$ may be carried out in the second term with the help of (\ref{A6}), and we find, 
\bea
\cD^{(4)} = 4 \zeta (4)^2 + \sum _{n_1 \not= 0} \, \, \sum _{n_2\not= 0, - n_1} { 1 \over n_1^2 \, n_2^2} 
\left ( { 2 \pi^2 \over 3 (n_1+n_2)^2} - { 6 \over (n_1+n_2)^4} \right )   
\eea
To sum over $n_2$, we make use again of (\ref{A6}) for the first term in the parentheses, and 
of the following summation formula for the second term in the parentheses, 
\bea
\sum _{n_2\not= 0, - n_1} { 1 \over n_2^2 (n_1+n_2)^4} = { \pi^4 \over 45 n_1^2} 
+ { 4 \pi^2 \over 3 n_1^4} - {15 \over n_1^6}   
\eea
The sums over $n_1$ may be performed in terms of ordinary $\zeta$-values, and we find,
\bea
\label{d44res}
\cD^{(4)} = 4 \zeta (4)^2 +{28 \pi^4 \over 45} \zeta (4) - 24 \pi^2 \zeta (6) +18 \zeta (8) = { \pi^8 \over 945}   
\eea
which is the leading asymptotic term in the expansion of $\cD$ in (\ref{3d4}).

\section{Harmonicity of $\cF_4$, $\cF_5$, and $\cF_6$.}
\setcounter{equation}{0}
\label{appB}

In this appendix, we shall collect and simplify the contributions to the coefficients $\cF_k$ in 
(\ref{4a2}) for $k=4,5,6$, arising from $\cD$, $\cC$, $\cE_2^2$ and $\cE_4$.  
We shall show that all contributions combine into a purely harmonic result. We shall calculate 
the functions $\f_k(q)$, and prove their conjugation properties, by inspection.  
The analysis of these contributions is considerably more complicated than the analysis for $k=0,1,2,3,7$ 
considered in the body of the text.

\subsection{Vanishing of non-harmonic terms in  $\cF _4$}
\label{secB1}

We begin by collecting all contributions to $\cF_4$ which are not {\sl manifestly harmonic}, and denote 
the result by $\cF_4^{{\rm nh}}$. It is given as follows,\footnote{Here, and in the following formulas, the prime 
superscript on the summation symbol indicates that the term with $m_r=0$ is omitted from the sum.}
\bea
\label{B1}
\cF_4 ^{{\rm nh}} & = & 
- \sum _{ m }'   { 1 \over m^4} \, { 1 + q^m \bar q^m \over |1-q^m|^2}
+ \sum _{m_1+m_2 \not= 0}'  {1 \over m_1^2 m_2^2 } \Re \left ( { 1+q^{m_1} \over 1 - q^{m_1}} \right )
\Re \left ( { 1+q^{m_2} \over 1 - q^{m_2}} \right )
\no \\ &&
+ \sum _{m_1 + m_2 \not= 0}'  { 6 \over m_1^2 m_2^2} 
\, { q^{m_1} \bar q^{m_2} \over (1-q^{m_1})^2 (1-\bar q^{m_2})^2} 
+  \sum _{m}' { 6 \over m^4} { q^m \bar q^m \over |1-q^m|^4}
\no \\ &&
- \sum _{m_1 , m_2 }'  { 6 \over m_1^2 m_2^2 }  { q^{m_1} \bar q^{m_2} \over (1-q^{m_1})^2 (1-\bar q^{m_2})^2} 
\eea
The first term arises from $\cD^{(2)}$, including the combinatorial factor of 6; 
the second term arises from $\cD^{(1)}$, including the combinatorial factor of 4, and
a factor of 3 to account for the sum over three equal terms in $\cD^{(1)}$; 
the third term arises from the non-generic case with a single independent pair vanishing in 
$\cD^{(0)}$, including the combinatorial factor of 3; 
the fourth term arises from the non-generic case with a two independent pairs vanishing in 
$\cD^{(0)}$, including the combinatorial factor of 3; and the fifth term arises from $- 3 \cE_2^2$.

\sm

$\bullet$ The last three terms in (\ref{B1}) manifestly cancel one another.

\sm

$\bullet$ To show the absence of non-harmonic terms in the first two terms, we use the identity, 
\bea
\label{B2}
\sum _m ' { 1 \over m^2} \Re \left ( { 1+q^m \over 1 - q^m} \right ) =0
\eea
which follows from the fact that  the summand is odd in $m$. Using (\ref{B2}), the second term of (\ref{B1}) 
is expressed as follows,  
\bea
\sum _{m_1+m_2 \not= 0}'  {1 \over m_1^2 m_2^2 } \Re \left ( { 1+q^{m_1} \over 1 - q^{m_1}} \right )
\Re \left ( { 1+q^{m_2} \over 1 - q^{m_2}} \right )
= 
 \sum _{m}' {1 \over m^4 } \left [ \Re  \left ( { 1+q^{m} \over 1 - q^{m}} \right ) \right ]^2
\eea
which may be decomposed into harmonic and non-harmonic sums as follows, 
\bea
\label{B4}
\sum _m' {1 \over m^4 } \left [ \Re \left ( { 1+q^m \over 1 - q^m} \right ) \right ]^2
= 
\sum _{m}' {1 \over 4 m^4 } \left (   { (1+q^{m})^2 \over (1 - q^{m} )^2} + { (1+\bar q^{m})^2 \over (1 - \bar q^{m} )^2} + 
2  { |1+q^{m}|^2 \over |1 - q^m |^2} \right )
\eea

\sm

$\bullet$ The first term of (\ref{B1}) can be usefully re-expressed with the help of the 
following rearrangement of the numerator,
\bea
1+q^m \bar q^m =  \half |1-q^m|^2 + \half |1+q^m|^2 
\eea
so that the sum becomes, 
\bea
\label{B6}
- \sum _{ m }'   { 1 \over m^4} \, { 1 + q^m \bar q^m \over |1-q^m|^2}
= 
- \sum _{ m }'   { 1 \over 2 m^4} 
- \sum _{ m }'   { 1 \over 2 m^4} \, { |1+q^m|^2 \over |1-q^m|^2}
\eea
The second term on the right side of (\ref{B6}) cancels the last term in the sum on the right side of (\ref{B4}).
The remaining terms, namely the first term on the right side of (\ref{B6}), and the first two terms in the sum  
on the right side of (\ref{B4}) are manifestly harmonic, and will need to be retained to compute $\f_4$.
Therefore   $\cF_4$ is harmonic.

\subsection{Calculation and properties of $\f_4(q)$}
\label{secB2}

Collecting all harmonic contributions to $\cF_4$ is most easily done by regrouping the terms that constitute $\f_4(q)$, 
and we find, 
\bea
\label{B7}
\f_4 (q) & = &
\sum _{m_1,m_2,m_3, m_4} ' {  \delta _{m,0} \over m_1 m_2 m_3 m_4} \, 
{ 1 \over (1-q^{m_1}) (1-q^{m_2}) (1-q^{m_3}) (1-q^{m_4}) }
\no \\ &&
- \sum _{m_1, m_2, m_3} ' { 12 \, \delta _{m,0}  \over  m_1^2 m_2 m_3} 
\, { 1+q^{m_1} \over (1-q^{m_1})^2 (1-q^{m_2}) (1-q^{m_3})} 
\no \\ && 
- \sum _{m_1, m_2} ' { 3 \over m_1 ^2 m_2^2} { q^{m_1}q^{m_2} \over (1-q^{m_1})^2 (1-q^{m_2})^2} 
+  \sum _m ' { 18 \over m^4} \, { q^{2m} \over (1-q^m)^4}
\eea
The first term on the right side arises from  $\cD^{(0)}$; 
the second term from $\cC^{(0)}$; the third term from $- 3 \cE_2^2$; and the fourth term arises
from combining the contributions from $ 18 \cE_4$,  $-48 \cC^{(2)}$, 
the harmonic terms in (\ref{B4}), and the constant term in (\ref{B6}).

\sm

It may be readily verified, by inspection term by term of (\ref{B7}),  that $\f_4(q)$ obeys the conjugation 
properties of (\ref{1a10}) for $k=4$. In particular, its Taylor series in powers of $q$ has real coefficients.

\subsection{Vanishing of non-harmonic terms in $\cF _5$}
\label{secB3}

We begin by collecting all contributions to $\cF_5$ which are {\sl not manifestly harmonic}, 
and denote the result by $\cF_5^{{\rm nh}}$. Its expression may be organized as follows,  
\bea
\label{B8}
\cF_5 ^{{\rm nh}} & = & 
\sum _m ' { 2 \over m^5} \, { 1 - q^m \bar q ^m \over |1-q^m|^2} + f_5 ^{(0)} + f_5 ^{(1)} + f_5 ^{(2)} 
- \sum _{m_1, m_2}' {3 \over  m_1^3 m_2^2} \Re \left ( {(1+q^{m_1} ) \, \bar q^{m_2} \over (1-q^{m_1} ) (1-\bar q^{m_2})^2} \right )
\no \\ &&
+ \sum _m ' { 6 \over  m^5} \, { (q^m + \bar q^m )(1-q^m\bar q^m) \over |1-q^m|^4}
- \sum _m ' { 6 \over  m^5} \, { 1-q^m \bar q^m \over |1-q^m|^2}
+ \sum _m ' { 1 \over  m^5} \, { 1-q^m \bar q^m \over |1-q^m|^2}
\no \\ &&
-\sum _{m_1+m_2\not=0} ' { 12  \over m_1^3 m_2^2} \, \Re \left ( { q^{m_1} (1+\bar q^{m_2}) 
\over (1-q^{m_1})^2 (1-\bar q^{m_2})} \right )
\eea
where we have used the following abbreviations, 
\bea
f_5^{(0)} &=&
- \sum _{m_r} ' { 3 \, \delta _{m,0} \over m_1 m_2 m_3^3} \Re \left ( { q^{m_1+m_2} + \bar q ^{m_3} 
\over (1-q^{m_1})(1-q^{m_2}) (1-\bar q^{m_3})} \right )
\no \\
f_5 ^{(1)} & = & 
- \sum _{m_r} ' { 3\, \delta _{m,0} (1-\delta _{m_1+m_2,0} ) \over 2 m_1 m_2 m_3 m_4 (m_1+m_2)} \,
\Re \left ( { q^{m_1+m_2} - \bar q^{m_3+m_4} \over (1-q^{m_1}) (1-q^{m_2}) (1-\bar q^{m_3}) (1-\bar q^{m_4})} \right )
\no \\
f_5 ^{(2)} & = &
- \sum _{m_r} ' { 2\,  \delta _{m,0}  \over m_1^2 m_2 m_3 m_4} \, 
\Re \left ( { q^{m_1} - \bar q^{m_2+m_3+m_4} \over (1-q^{m_1}) (1-\bar q^{m_2}) (1-\bar q^{m_3}) (1-\bar q^{m_4})} \right ) 
\eea
The first term in (\ref{B8}) arises from $6 \cD_4^{(2)}$, the second from $4 \cD_4 ^{(1)}$, the third and fourth from $\cD_4^{(0)}$, the fifth from $-3 \cE_2^2$, the sixth and seventh from $-48 \cC^{(2)}$, the eighth from $-24 \cC^{(1)}$, and the ninth from $-24 \cC^{(0)}$. We have the following simplifications.

\sm

$\bullet$ The first term on the second line of (\ref{B8})  cancels the term on the third line,  using (\ref{B2}).

\sm

$\bullet$ To simplify $f_5^{(0)}$, we use the following identity for the numerator in the sum, 
\bea
q^{m_1+m_2} + {\bar q}^{m_3} 
= 
{1 \over 4} \sum _{\sigma , \sigma ' = \pm 1}  
\sigma' (1+\sigma \, q^{m_1}) (1+ \sigma \sigma ' \, q^{m_2}) (1+\sigma' \, {\bar q}^{m_3})
\eea
as a result of which we obtain, 
\bea
f_5 ^{(0)} & = & \hat f_5^{(0)} 
- \sum _{m_r} ^\prime  { 3 \, \delta _{m,0} \over 4 m_1 m_2 m_3^3} \, \Re \left ( { 1+\bar q ^{m_3} \over 1- \bar q ^{m_3}} \right ) 
+ \sum _{m_r} ^\prime  { 3 \delta _{m,0} \over 2 m_1 m_2 m_3^3} \, \Re \left ( { 1+q^{m_1} \over 1- q^{m_1}} \right ) 
\no \\ 
\hat f_5 ^{(0)} & = & 
-  \sum _{m_r} ^\prime  { 3 \, \delta _{m,0} \over 4 m_1 m_2 m_3^3} \, 
\Re \left ( { (1+ q ^{m_1}) (1+ q ^{m_2}) (1+\bar q ^{m_3}) \over (1- q ^{m_1}) (1- q ^{m_2}) (1- \bar q ^{m_3})} \right ) 
\eea

\sm

$\bullet$ To simplify $f_5^{(1)}$ we use the following decomposition formula,
\bea
q^{m_1+m_2} - \bar q^{m_3+m_4} & = & 
+ {1 \over 4} \sum _{\sigma = \pm 1} (1 + q^{m_1+m_2}) (1- \sigma \, \bar q^{m_3} )(1+ \sigma \, \bar q^{m_4})
\no \\ &&
- {1 \over 4} \sum _{\sigma = \pm 1}  (1 + \sigma \, q^{m_1}) (1- \sigma \, q^{m_2})  (1+\bar q^{m_3+m_4} )
\eea
Symmetry of the sum in $f_5^{(1)}$ under permutations  $m_1 \leftrightarrow m_2$ and $m_3 \leftrightarrow m_4$, 
as well as under complex conjugation combined with the reversal of signs of all $m_r$, implies that all four terms 
above produce equal contributions  to $f_5^{(1)}$. As a result, the simplified $f_5^{(1)}$  is given as follows,
\bea
f_5 ^{(1)} = 
- \sum _{m_r} ' { 3 \, \delta _{m,0} (1-\delta _{m_1+m_2,0} ) \over 2 m_1 m_2 m_3 m_4 (m_1+m_2)} \,
\Re \left ( { (1+q^{m_1+m_2}) (1+\bar q^{m_3})  \over (1-q^{m_1}) (1-q^{m_2}) (1-\bar q^{m_3}) } \right )
\eea
Further simplification is achieved by using the following decomposition,
\bea
1+q^{m_1+m_2} = \half (1-q^{m_1}) (1-q^{m_2}) + \half (1+q^{m_1}) (1+q^{m_2})
\eea
as a result of which we obtain, 
\bea
f_5 ^{(1)} & = & \hat f_5 ^{(1)} 
- \sum _{m_r} ' { 3 \, \delta _{m,0} (1-\delta _{m_1+m_2,0} ) \over 4 m_1 m_2 m_3 m_4 (m_1+m_2)} \,
\Re \left ( {  1+ \bar q^{m_3}  \over  1- \bar q^{m_3} } \right )
\\ 
\hat f_5 ^{(1)}  & = & 
- \sum _{m_r} ' { 3 \, \delta _{m,0} (1-\delta _{m_1+m_2,0} ) \over 4 m_1 m_2 m_3 m_4 (m_1+m_2)} \,
\Re \left ( { (1+q^{m_1})(1+q^{m_2}) (1+\bar q ^{m_3})  \over (1-q^{m_1}) (1-q^{m_2}) (1-\bar q ^{m_3}) } \right )
\no
\eea
Note that the second  term of the first of these equations is harmonic.

\sm

$\bullet$ To simplify $f_5^{(2)}$, we use the following decomposition formula,
\bea
q^{m_1} - \bar q^{m_2+m_3+m_4} =
\half \sum _{\sigma = \pm 1} (1+ \sigma \, q^{m_1} ) ( 1 - \sigma \, \bar q ^{m_2+m_3+m_4})
\eea
The contribution from the $\sigma =-1$ term above is harmonic, while the contribution from the 
$\sigma =+1$ term may be further simplified by using the following decomposition formula,
\bea
1- \bar q^{m_2+m_3+m_4} = 
{1 \over 4} \sum _{\sigma , \sigma ' =\pm 1} 
 (1+\sigma \, \bar q ^{m_2}) (1+\sigma ' \, \bar q^{m_3}) (1- \sigma \sigma ' \, \bar q ^{m_4})
\eea
Hence we find the simplified formula,
\bea
f_5 ^{(2)} & = & \hat f_5 ^{(2)} 
+ \sum _{m_r} ' {  \delta _{m,0}  \over m_1^2 m_2 m_3 m_4} \, 
\Re \left ( { 1+ \bar q ^{m_2+m_3+m_4} \over (1-\bar q^{m_2}) (1-\bar q^{m_3}) (1-\bar q ^{m_4})} \right ) 
\no \\ &&
- \sum _{m_r} ' {    \delta _{m,0}  \over 4 m_1^2 m_2 m_3 m_4} \, 
\Re \left ( { 1+q^{m_1}  \over 1-q^{m_1}  } \right ) 
\no \\ 
\hat f_5 ^{(2)} & = &
- \sum _{m_r} ' {  3 \,  \delta _{m,0}  \over 4 m_1 m_2 m_3^2 m_4} \, 
\Re \left ( { (1+q^{m_1}) (1+q ^{m_2}) (1+\bar q ^{m_3})  \over (1-q^{m_1}) (1-q ^{m_2}) (1-\bar q ^{m_3}) } \right ) 
\eea
where in writing $\hat f_5^{(2)}$ we have taken the liberty of permuting the indices $m_1$ and $m_3$ and taking the complex conjugate of the expression under the reality sign, for later convenience.
Note that the second and third terms of the first of these equations are harmonic.

\sm

$\bullet$ To combine $\hat f_5^{(1)}$ and $\hat f_5 ^{(2)}$ we make use of the following identity,
\bea
- { 3 \, \delta _{m,0} (1-\delta _{m_1+m_2,0} ) \over 4 m_1 m_2 m_3 m_4 (m_1+m_2)}
- {  3 \,  \delta _{m,0}  \, (1-\delta _{m_1+m_2,0})  \over 4 m_1 m_2 m_3^2 m_4}
= 
 { 3 \, \delta _{m,0} (1-\delta _{m_1+m_2,0} ) \over 4 m_1 m_2 m_3 ^2 (m_1+m_2)}
\eea
We may  carry out the sum over $m_4$ explicitly, 
and express the result in the following form,
\bea
\label{B18}
\hat f_5^{(1)} + \hat f_5 ^{(2)}
& = &
\hat f_5 ^{(3)} + \sum _{m_1, m_3} ' {  3   \over 4 m_1^2  m_3^3 } \, 
\Re \left ( { (1+q^{m_1})^2  (1+\bar q ^{m_3})    \over (1-q^{m_1})^2  (1-\bar q ^{m_3})  } \right ) 
\no \\
\hat f_5 ^{(3)} & = &
 \sum _{m_r} '  {  3 \, \delta _{m,0}  \over 4 m_1 m_2  m_3^3} \,  \, 
\Re \left ( { (1+q^{m_1}) (1+q^{m_2}) (1+\bar q ^{m_3})    \over (1-q^{m_1}) (1-q^{m_2})  (1-\bar q ^{m_3})  } \right ) 
\eea
In expressing $\hat f_5^{(3)}$ on the second line above, we have carried out the sum over $m_4$ to eliminate $\delta _{m,0}$, which imposes the condition  $m_1+m_2+m_3 \not= 0$ on the remaining sum. 

$\bullet$ It is clear by inspection that $\hat f_5^{(0)} + \hat f_5 ^{(3)}=0$. The remaining contributions to $\cF_5$ which are not yet manifestly harmonic arise from the first, fifth, seventh, and eighth terms in (\ref{B8}) (the sixth and ninth terms cancel one another, as shown earlier), as well as from the second term on the right side of the first equation in (\ref{B18}). Assembling those contributions gives,
\bea
&& \sum _{m_1, m_2} ' {  3   \over 4 m_1^2  m_2^3 } \, 
\Re \left ( { (1+q^{m_1})^2  (1+{\bar q}^{m_2})    \over (1-q^{m_1})^2  (1-{\bar q}^{m_2})  } \right ) 
\no \\ &&
- \sum _{m_1, m_2}' {3 \over  m_1^2 m_2^3} 
\Re \left ( { \, q^{m_1} \, (1+{\bar q}^{m_2} )  \over  (1-q^{m_1})^2 (1-{\bar q}^{m_2} )} \right )
- \sum _m ' { 3 \over  m^5} \, { (1-q^m{\bar q}^m) \over (1-q^m) (1-{\bar q}^m)}
\eea
it is easy to see that the sum is in fact harmonic as well. 

\sm

In summary, we have shown that $\cF_5$ is purely harmonic.

\subsection{Calculation and properties of $\f_5(q)$}
\label{secB4}

Collecting all harmonic contributions to $\cF_5$ is most easily done by regrouping the terms that constitute $\f_5(q)$, and we find,
\bea
\label{B22}
\f_5 (q) & = & 
\sum _{m_1,m_2,m_3, m_4} ' {  \delta _{m,0}  \over 2 m_1^2 m_2 m_3 m_4} \, 
 { 1+ q^{m_2+m_3+m_4} \over (1-q^{m_2}) (1-q^{m_3}) (1-q^{m_4})}  
 \no \\ &&
 - \sum _{m_1,m_2,m_3} '  { ( 18 m_1 + 12 m_2)  \delta _{m,0} \over m_1^3 m_2m_3^2} \, 
{ q^{m_1} ( 1+q^{m_2}) \over (1-q^{m_1})^2 (1-q^{m_2})} 
\, 
+ \sum _{m_1} ' { 9 \over m_1^5} \, { q^{m_1} (1+q^{m_1}) \over (1-q^{m_1})^3} 
\no \\ &&
- \sum _{m_1, m_2}' {3 \over 2 m_1^3 m_2^2} \, { (1+q^{m_1} ) \, q^{m_2} \over (1-q^{m_1} ) (1-q^{m_2})^2} 
+ \sum _{m_1} ' \phi (m_1) \, { 1+q^{m_1} \over 1- q^{m_1}} 
\eea
The coefficient $\phi(m_1)$  in the last term is given by the following multiple sums, 
\bea
\phi(m_1) & = &
-{ 3 \over  2 m_1^5} 
+ \sum _{m_2} ' {  3   \over 8 m_1^3  m_2^2 } 
- \sum _{m_2, m_3} ' { 3 \, \delta _{m,0} \over 8 m_1^3 m_2 m_3}  
+ \sum _{m_2,m_3} ' { 3 \, \delta _{m,0} \over 4 m_1 m_2 m_3^3}  
\no \\ &&
- \sum _{m_2,m_3,m_4} ' {  \delta _{m,0}  \over 8 m_1^2 m_2 m_3 m_4}  
 - \sum _{m_2,m_3,m_4} ' { 3 \, \delta _{m,0} (1-\delta _{m_3+m_4,0} ) \over 8 m_1 m_2  m_3 m_4 (m_3+m_4)} 
 \eea
It may be readily verified, by inspection of (\ref{B22}),  that $\f_5(q)$ obeys the conjugation 
properties of (\ref{1a10}) for $k=5$. In particular, its Taylor series in powers of $q$ has only real coefficients.

\subsection{Vanishing of non-harmonic terms in $\cF _6$}
\label{secB5}

We begin by collecting all contributions to $\cF_6$ which are {\sl not manifestly harmonic}, 
and denote the result by $\cF^{{\rm nh}}_5$. Its expression may be organized as follows, 
\bea
\cF_6 ^{{\rm nh}} & = &
\sum _{m}' { 33 \over 4 m^6} \, { 1+ q^m \bar q ^m \over (1-q^m) (1-\bar q^m)}
\no \\ &&
+ \sum _{m_r} ' { 3 \, \delta _{m,0} \over 2 m_1 m_2 m_3^4} 
\Re \left ( { q^{m_1+m_2} -\bar q^{m_3} \over (1-q^{m_1}) (1-q^{m_2}) (1 - \bar q^{m_3})} \right )
\no \\ &&
- \sum _{m_1, m_2}' { 3 \over 8 m_1^3 m_2^3} 
\Re \left ( { (1+q^{m_1})(1+\bar q^{m_2})  \over (1- q^{m_1})(1-\bar q^{m_2}) } \right )
\no \\ &&
- \sum _{m_1+m_2\not= 0} ' { 3 \over 2 m_1^3 m_2^3} \, \Re \left ( 
{ q^{m_1}  + \bar q^{m_2} \over (1-q^{m_1})(1-\bar q^{m_2})} \right )
\no \\ &&
+ \sum _{m_1+m_2\not= 0} ' { 3 (m_1-3m_2)  \over 2 m_1^4 m_2^3} \, \Re \left ( 
{ (1+q^{m_1} ) (1+\bar q^{m_2})  \over (1-q^{m_1})(1-\bar q^{m_2})} \right )
\eea
The first line arises from combining the contributions from $6\cD_4^{(2)}$, $-48\cC^{(2)}$, and $-24\cC^{(1)}$;
the second line arises from $4\cD^{(1)}$, the third from $-3\cE_2 ^2$;
the fourth from the fourth line in $-24\cC^{(0)}$;
and the fifth from the last two lines in $-24\cC^{(0)}$  (which contribute equally).

\sm

$\bullet$ The sums in the second and fourth lines may be simplified by using the following identities for their numerator, 
\bea
q^{m_1+m_2}  - \bar q^{m_3} & = & 
{1 \over 4} \sum _{\sigma , \sigma ' = \pm 1} 
\sigma ' (1+ \sigma \, q^{m_1}) (1 + \sigma \sigma ' \, q^{m_2}) (1- \sigma ' \, \bar q^{m_3}) 
\no \\
q^{m_1} + \bar q^{m_2} & = & \half (1+q^{m_1} ) (1+\bar q^{m_2} ) - \half (1-q^{m_1} ) (1-\bar q^{m_2} )
\eea
The contribution with $\sigma '=+1$ in the first line is harmonic, as is the contribution from the second term in the second line above. The remaining not-manifestly-harmonic contributions are as follows,
\bea
\label{B25}
-{ 3 \over 8} f_6 
- \sum _m ' { 9 \over 8 m^6} \, { (1+q^m) (1+\bar q^m) \over (1-q^m) (1-\bar q^m)} 
+ \sum _{m_1,m_2} ' { 3  \over 8 m_1^3 m_2^3} \, \Re \left ( 
{ (1+q^{m_1} ) (1+\bar q^{m_2})  \over (1-q^{m_1})(1-\bar q^{m_2})} \right )
\eea
where we have introduced the notation, 
\bea
f_6 =  \sum _{m_r} ' \! {   \delta _{m,0} \over   m_1 m_2 m_3^4} 
  {(1+q^{m_1}) (1+ \bar q^{m_3})  \over (1-q^{m_1})  (1 - \bar q^{m_3})} 
  + \sum _{m_r} ' \! {  \delta _{m,0} \over  m_1 m_2 m_3^4} 
 {(1+q^{m_3}) (1+ \bar q^{m_1})  \over (1-q^{m_3})  (1 - \bar q^{m_1})}
\eea
In the second sum, we relabel the indices by permuting $m_1$ and $m_3$, which makes the dependence 
on $q$ and $\bar q$ the same in both terms, and gives,
\bea
f_6 =   \sum _{m_1,m_2, m_3} ' 
\left ( {  \delta _{m,0} \over  m_1 m_2 m_3^4} + {  \delta _{m,0} \over  m_1^4 m_2 m_3 } \right ) 
  {(1+q^{m_1}) (1+ \bar q^{m_3})  \over (1-q^{m_1})  (1 - \bar q^{m_3})} 
\eea
We now use the following rearrangement formula, 
\bea
{  \delta _{m,0} \over  m_1 m_2 m_3^4} + {  \delta _{m,0} \over  m_1^4 m_2 m_3 }
= - { \delta _{m,0} \over  m_1^4 m_3^4} \, ( m_1^2 -  m_1 m_3 + m_3^2)
\eea
As a result, $f_6 $ becomes, 
\bea
f_6 = 
 \sum _{m_1,m_2, m_3} ' 
\left (  {  \delta _{m,0} \over  m_1^3 m_3^3} -  {  2 \, \delta _{m,0}  \over  m_1^2 m_3^4}  \right ) 
\Re \left (   {(1+q^{m_1}) (1+ \bar q ^{m_3})  \over (1-q^{m_1})  (1 - \bar q ^{m_3})} \right )
\eea
Using (\ref{B2}) in the sum for the last term in the parentheses, we obtain, 
\bea
\label{B30}
f_6 = 
\sum _{m_1,m_3} ' 
 { 1 \over  m_1^3 m_3^3} 
\Re \left (   {(1+q^{m_1}) (1+ \bar q ^{m_3})  \over (1-q^{m_1})  (1 - \bar q ^{m_3})} \right )
- \sum _{m} ' 
 {  3   \over  m^6}    {(1+q^{m}) (1+ \bar q ^{m})  \over (1-q^{m})  (1 - \bar q ^{m})}   
\eea
The expression (\ref{B25}) is seen to vanish by substituting  the expression for $f_6$ derived in (\ref{B30}).

\sm

This concludes the proof that $\cF_6$ is purely harmonic.

\subsection{Calculation and properties of $\f_6(q)$}
\label{secB6}

Collecting all harmonic contributions to $\cF_6$ is most easily done by regrouping the terms that 
constitute $\f_6(q)$, and we find,
\bea
\label{B31}
\f_6 & = & 
  - \sum _{m} ' { 9 \over 16 m^6 } \, { (1+q^m)^2 \over ( 1- q^m)^2} 
+ \sum _{m_1,m_2}' { 3  \over 16 m_1^3 m_2^3} \, 
{ (1+q^{m_1})(1+q^{m_2})  \over (1- q^{m_1})(1-q^{m_2}) } 
\no \\ &&
+ \sum _{m_1+m_2 \not= 0}' { 9 \over 8 m_1^2 m_2^2 (m_1+m_2)^2}
\, { (1+q^{m_1} ) (1 + q^{m_2}) \over (1-q^{m_1} ) (1 - q^{m_2})} 
\no \\ && 
+  \sum _{m_1+m_2 \not= 0} ' { 3   \over 16 m_1 m_2 (m_1+m_2)^4} 
  { (1+q^{m_1}) (1 + q^{m_2}) \over (1-q^{m_1}) (1-q^{m_2}) } 
  \eea
It may be readily verified, by inspection of (\ref{B31}),  that $\f_6(q)$ obeys the conjugation 
properties of (\ref{1a10}) for $k=6$. In particular, its Taylor series in powers of $q$ has only real coefficients.

\section{Details of the analysis of $K(\tau,y)$} 
\setcounter{equation}{0}
\label{appC}

In this Appendix,\footnote{The analysis in this appendix was suggested by Stephen Miller.}  
we shall show that $K(\tau, y)$, which has been defined in (\ref{4a7}) and is of the form (\ref{4a8})
with $K_4(\tau) , K_5 (\tau), K_6 (\tau)$ holomorphic in $\tau $ in the upper half plane $\mH$, 
and which is purely imaginary for all $\tau$ and $y$, 
must be of the form given in (\ref{Kres}), with $K_4,K_5,K_6$ polynomials in $\tau$ of degree one, 
two, and three respectively. This is a crucial auxiliary result, which is used in section \ref{sec4} 
to prove the Theorem of this paper.

\sm

Note that $K(\tau,y)$ must vanish along the imaginary axis, as on the one hand it must be purely imaginary, yet on the other hand, its building blocks $H$ and $H^S$ are real there. If $K(\tau, y)$  had been a holomorphic function, then its vanishing along the imaginary axis would have implied its vanishing everywhere.  However, $K(\tau, y)$ is only {\sl almost holomorphic}, so the situation is more subtle, as we will see.

\sm

We will Taylor expand $y^2K(\tau,y)$ around an arbitrary point on the imaginary axis in the upper half 
$\tau$-plane $\mH$. 
It will be convenient to choose that point to be $\tau=i$, and to expand in the variable $\ttau$ (or in the variables 
$\tilde x, \tilde y$), defined by,
\bea
\tau = i + \ttau \hskip 1in \tau = \tilde x + i \tilde y, \quad \tilde x, \tilde y \in \RR
\eea
The expansion of the holomorphic functions $\psi _k(\tau)$ around $\tau=i$ is best organized in terms of their expansion in terms of (positive powers of) $q$, and we have, 
\bea
\psi_k(\tau) = \sum_{n=0}^\infty d_{k,n} \,  e^{2\pi i n \ttau}
\label{phigen}
\eea
where the coefficients $d_{k,n}$ are real. (Displacing the expansion point to another point on the imaginary 
axis will maintain reality of the coefficients $d_{k,n}$, but change their values.) 

\sm

We will now use the fact that the $S$ transformation $\tau \to -1/\tau$
leaves the point $\tau=i$ invariant, and takes the following form on $\tilde \tau$, 
 \bea
i + \ttau \to  -\frac{1}{i + \ttau}= { i \over 1- i \ttau} =  i \sum _{n=0} ^\infty  i^n \ttau^n 
 \label{tauinv}
 \eea 
 to write the expansion of  $ K(\tau,y) \equiv H(\tau,y) -  H^S(\tau,y)$  in powers of $\tx$ and $\ty$,
   \bea
y^2 K(\tau,y) = \sum_{r=0}^2 \sum_{n=0}^\infty  \ty^r \,c_{n,r}\, i^n\, (\tx+i\ty)^n 
\label{Kexpand}
\eea
where the coefficients $c_{n,r}$ are again real constants.   Expanding the $n$-th power gives,
  \bea
y^2 K(\tau,y) = \sum_{r=0}^2 \sum_{n=0}^\infty \sum_{l=0}^n c_{n,r}\, \left({n \atop l}\right)\,(-1)^l\, i^{n-l} \, \tx^{n-l} \, \ty^{l+r}  
\label{newK}
\eea
setting $n=a+l\ge 0$ and $l=b-r\ge 0$ we  have
\bea
y^2 K(\tau,y) = \sum_{a=0}^\infty   
\sum_{r=0}^2\sum_{b=r}^{\infty} c_{a+b-r,r}\, \left({a+b-r \atop b-r}\right)\,(-1)^{b-r}\, i^{a}\, \tx^a\, \ty^b
\label{abcoeff}
\eea
Since $K(\tau, y)$ must be imaginary for all $\tilde x$ and $\tilde y$, we conclude that all contributions with even index
$a$ must vanish identically as a function of $\tilde y$. The terms with $b=0$ and $b=1$ are special in that they receive contributions only from $r=0$ and $r=0,1$ respectively. The conditions for $b=0,1$, and all even $a$, are as follows,
\bea
c_{a,0}=0 
\hskip 1in 
 c_{a,1} = (a+1) c_{a+1,0}
\eea
The conditions for $b \geq 2$, and  all even $a$, are as follows,
\bea
c_{a+b,0}\, \left({a+b \atop b}\right)  - c_{a+b-1,1}\, \left({a+b-1 \atop b-1}\right) + c_{a+b-2,2}\, \left({a+b-2 \atop b-2}\right) =0
\eea
The equations for $0 \leq b \leq 5$ may be readily solved: they set all coefficients $c_{n,r}$ to zero, except for the following set, 
\bea
c_{1,0} \hskip 0.5in 
c_{3,0} \hskip 0.5in 
 c_{1,1} \hskip 0.5in  
 c_{2,1} = 3 c_{3,0}
 \hskip 0.5in  c_{1,2} = 2c_{3,0}
\label{nonzero}
\eea
whose values are left undetermined by the equations. The equations for $b \geq 6$ do not impose any further conditions on the coefficients $c_{n,r}$.  So we find that consistency with modular invariance places a  very restrictive condition on $K(\tau, y)$. 

\sm

Substituting this result into the expansion (\ref{abcoeff}) shows that $K(\tau,y)$ is given in terms of  a very small number of terms with odd powers of $\tilde x$,
\bea
 y^2 K(\tau,y) =i \tx \Big ( c_{1,0} + c_{1,1}\, \ty + c_{3,0}\, (\tx^2+\ty^2) \Big )
\label{econ}
\eea
Since the expression only has a finite number of powers of $\tilde x+ i \tilde y = \ttau$ we can return to the original $\tau$ coordinate by replacing $\ttau\to \tau -i$, $\ty\to y-1$, which results in
 \bea
 K(\tau,y) = A+ 2 i \, B\,  \tau + \frac{1}{y} (C + i\, A\, \tau - 3\, B\, \tau^2)+ \frac{1}{y^2} (i\, C\, \tau - i\, B\, \tau^3 )
\eea
where the coefficients $A=c_{11} + 2\, c_{30}$, $B=c_{3,0}$, $C=c_{1,0}-c_{1,1} - c_{3,0}$ are all real.
This result holds for $\tau $ in the upper half plane, and may be continued to the lower half plane 
using the second equation in  (\ref{4a7a}). The resulting $K$ in the lower half plane has reversed signs 
of $B$ and $C$, but the same sign of $A$.

\section{Deligne's proof of the vanishing of $\f_1,\f_2,\f_3$} 
\setcounter{equation}{0}
\label{appD}

In this appendix we present an alternative and  more direct proof, due to Deligne \cite{Deligne}, 
for the results obtained in section \ref{sec4}.  

\sm

We begin by deriving the following result. A function $F_N(\tau, \bar \tau)$ on the Poincar\'e upper half plane $\mH= \{ \tau=x+iy, ~ x, y \in \RR, ~y>0\}$, of the form,
\bea
\label{D1}
F_N (\tau, \bar \tau) = \sum _{k=0} ^N { 1 \over y^k} \, \Big ( A_k (q) + B_k (\bar q) \Big )
\eea
with $F_N$ invariant under the action of  $SL(2,\ZZ)$, and the functions $A_k(q)$ and $B_k(q)$ 
holomorphic in the unit disc $|q| < 1$ for all $0\leq k \leq N$, is necessarily a constant.  Here we continue to denote $q=e^{2\pi i \tau}$.  The function $F(\tau,\bar \tau)$ in (\ref{4a1}) is a special case  of such a function with $B_k(\bar q) = \overline{A_k(q)}$  and $N=2$ but the following argument holds for all $N$ and general $A_k,  B_k$.

\sm

The proof relies on the following two important properties. First, the function $F_N$ is in $L^2(\cM_1)$,
namely it is square integrable on $\cM_1= \mH /PSL(2,\ZZ)$ with the Poincar\'e measure $dx dy / y^2$. 
To see this, we use the $SL(2,\ZZ)$-invariance of $F_N$ to represent $\cM_1$ by the customary fundamental 
domain $\cM_1 = \{ \tau \in \mH, ~ |\tau|>1, ~ | x | < \half\}$. The $L^2$ property then follows from the fact that $y$ is bounded away from $0$ on $\cM_1$ and the fact that $A_k$ and $B_k$ are bounded on $\cM_1$. Second, $F_N$ satisfies the following differential equation, 
\bea
\label{D2}
\left ( \prod _{k=0}^N \left ( \Delta - k(k+1) \right ) \right ) F_N =0
\eea
where $\Delta $ denotes the Laplace operator $\Delta = 4 y^2 \p_\tau \p _{\bar \tau}$. This result may be established 
by observing that the function $(\Delta - N(N+1)) F_N$ is a function of the type $F_{N-1}$. Iterative application of the remaining factors $(\Delta - k(k+1)$ for $k=1, \cdots, N-1$ thus produces a harmonic function, of type $F_0$, which is annihilated by the factor for $k=0$.

\sm

Using the above results, we now proceed to solving the differential equation (\ref{D2}). The spectrum of $\Delta $ on $\cM_1$ is real and non-positive. It has a continuous part, with eigenvalues given by $-{1 \over 4} -s^2$ and associated eigenfunctions given by the Eisenstein series $E_{- \half + i s}$ for real $s$, and a discrete spectrum embedded in the continuous spectrum, except for the complementary series which here just contains eigenvalue 0 with corresponding constant eigenfunction \cite{Zagier:2008, Terras}. The values $k(k+1) $ for $k \in \NN$ are outside this spectrum, except for the value $k=0$ for which we have a constant eigenfunction. Hence the operators $(\Delta - k(k+1) )$ are all invertible for $k\not=0$ and the only solution to (\ref{B2})   left is the constant function for $k=0$.

\sm

Applying the above result to the case $N=2$,  we find that $F$ in (\ref{1a7}) must be constant, but this constant must vanish by inspection of the asymptotic behavior of $F$ as $\tau \to i \infty$.

\end{document}